%% file: main-generic.tex
\documentclass[a4paper]{article}

\usepackage[english]{babel}
\usepackage[utf8]{inputenc}
\usepackage{amsmath}
\usepackage{amsthm}

\usepackage{graphicx}
\usepackage[colorinlistoftodos]{todonotes}

 \usepackage[numbers,super]{natbib}%

\bibpunct{[}{]}{,}{n}{}{}%
\setcitestyle{number,super,open={[},close={]}}
 
\usepackage[margin=0.5in]{geometry}
\usepackage{authblk}
\usepackage{hyperref}

\title{A simple thermodynamic framework for heat-conducting flows of mixtures of two interacting fluids}

\date{}                     

\author[1]{Josef M\'{a}lek}
\author[1]{Ond\v{r}ej Sou\v{c}ek}
\affil[1]{Charles University, Faculty of Mathematics and Physics, Mathematical Institute, Sokolovsk\'{a} 83, 186 75 Prague 8, Czech Republic}

\date{\today}

\include{packages}

\begin{document}
\include{macros-generic}

\newtheorem{remark}{Remark}

\maketitle

\begin{abstract}
Within the theory of interacting continua, we develop a model for a heat conducting mixture of two interacting fluids described in terms of the densities and the velocities for each fluid and the temperature field for the mixture as a whole. We use a general thermodynamic framework that determines the response of the material from the knowledge of two pieces of information, namely how the material stores the energy and how the entropy is produced. This information is expressed in the form of the constitutive equations for two scalars: the Helmholtz free energy and the entropy production. Additionally, we follow the goal to determine the response of a mixture from a small (minimal) set of material parameters, including shear viscosity, bulk viscosity and heat conductivity associated with the mixture as a whole and the drag coefficient connected with the interaction force between the constituents. The same thermodynamic approach is used to obtain the model when the mixture as a whole responses as an incompressible material. For both the compressible and incompressible mixtures, we investigate three variants stemming from different definitions of the (averaged) velocity associated with the mixture as a whole. We also address the issue of identification of boundary conditions for the individual constituents from the standard boundary conditions formulated in terms of the quantities associated with the mixture as a whole.\end{abstract}

\section{Introduction}

The aim of this study is to develop a simple and, in the sense specified below, minimalist model capable of describing heat-conducting flows of mixtures consisting of two interacting fluids (liquids or gases).  We stem from the basic principles of the {\it theory of interacting continua} established by Truesdell \cite{Truesdell-1962}, 
see also the survey articles by 
M\"{u}ller \cite{Muller-1968}, Atkin and Craine\citep{Atkin-Craine-1976}, Bowen \cite{bowen.rm:continuum}, Bothe and Dreyer~\cite{bothe-2015} and the books by Samoh\'{y}l\cite{Samohyl-1987}, Rajagopal and Tao\cite{Rajagopal-and-Tao-1995}, Hutter and J\"{o}hnk \cite{hutter.k.johnk.k:continuum}, or Peka\v{r} and Samoh\'{y}l \cite{Pekar-2014}. This theory is based on the assumption that the all constituents coexist at each point of the current configuration occupied by the mixture. The governing equations then express the balance equations for mass, linear and angular momenta and energy associated with each constituent and are completed by the formulation of the second law of thermodynamics (balance equation for the entropy with the requirement that the entropy production is non-negative) associated with the mixture as a whole (whole-mixture in short). 

This general setting can be further simplified by the requirement that the temperatures of the individual constituents coincide which allows one to consider the balance of energy for the whole mixture expressed however in the form of the sum of the balance equations for energy of individual constituents. This assumption will be adopted in our study. 
We also restrict ourselves to a \textit{binary mixture}, i.e. to a mixture consisting of two constituents (\textit{two fluids}). Despite these simplifications, the set of governing equations contains quantities such as the energy flux, the entropy flux, the entropy production, the Cauchy stresses associated with each fluid and the mass and momenta interaction terms, that all enter into the constitutive equations characterizing the response of the whole mixture. These constitutive equations (that can be in the form of algebraic or evolutionary partial differential equations) are needed in order to obtain a closed system of the equations describing processes of the considered mixtures. The goal of this study is to develop \textit{the simplest} possible \textit{thermodynamic framework} leading to the forms of these constitutive equations, under the assumptions stated above. 

The requirement of \textit{simplicity} is connected with the objective to provide a framework in which the complete model is obtained from the knowledge of the material properties (shear viscosity, bulk viscosity and heat conductivity) associated with the whole mixture and where the only interaction mechanism between the constituents (apart from possible mass conversion due to chemical reactions or phase change) is the drag force. This requirement (minimalist regarding the number of material coefficients) is motivated by the fact that these are exactly the parameters that can be experimentally measured. In this aspect we follow the article by M\'{a}lek and Rajagopal\cite{malek-2008}. The framework developed there carries on a thermodynamic approach developed in Rajagopal and Srinivasa~\cite{Rajagopal-Srinivasa-2004}. This approach is based on the idea that the complete response of the material can be determined from the specification of the constitutive equations for two scalar quantities, namely the specific entropy (or any of related thermodynamic potentials: internal energy, Helmholtz free energy, enthalpy or Gibbs potential) and the entropy production. From the knowledge of these two scalar quantities, one can derive the complete model involving, in particular, the constitutive equation for the Cauchy stress and the energy flux. This thermodynamic approach has been successfully applied, in many areas, to the development of the models that are capable of describing complicated responses of materials whereas the resulting model is automatically consistent with the laws of thermodynamics (we refer to Rajagopal and Srinivasa~\cite{Rajagopal-Srinivasa-2004}, M\'{a}lek and Rajagopal~\cite{malek-2008}, 
Rajagopal and Srinivasa~\cite{krrsri2011}, Málek, Rajagopal, Tůma~\cite{mkrrkt2015, mkrrkt2018}, Kratochv\'\i{}l, Málek and Minakowski~\cite{KrMaMi2016}, Cichra and Pr\r{u}\v{s}a~\cite{cipr2020},  and a survey paper M\'{a}lek and Pr\r{u}\v{s}a~\cite{Malek2016} for further details and references). 

Although we are following the principle idea developed in M\'{a}lek and Rajagopal~\cite{malek-2008}, here we are able to overcome the following three shortcomings of their study. First, in M\'{a}lek and Rajagopal~\cite{malek-2008}, the functional form describing the mechanism of energy storage was postulated to be the same for each constituent. Second, the study~\cite{malek-2008} was restricted to isothermal processes. Third, in thermodynamic equilibrium, the model developed in~ \cite{malek-2008} was not compatible with the mixture of ideal gases. This study aims at removing all these deficiencies. 

Doing so, we also revisit the definitions of the (averaged) velocity for the whole mixture. The velocity of the mixture can be defined by means of the velocities of the individual constituents weighted, for example, by mass concentrations, molar concentrations or volume fractions. It is known that each of the mixture velocity definitions has its advantages. For instance, the most traditional mass-weighted definition of (barycentric) mixture velocity admits particularly simple form of the balance equations formulated for the whole mixture. The volume-weighted mixture velocity allows for a divergence-free formulation of the momentum balance for a large class of so-called quasi-incompressible materials, which facilitates rigorous mathematical analysis, see Abels et al. \cite{abels2012,abels2013}. For a comprehensive comparison of the mass and volume weighted forms of balance equations, see \v{R}eho\v{r}~\cite{rehor-2018}. Interestingly, it appears that the choice of molar-weighted whole-mixture velocity directly leads to a model that meets all three above stated requirements we wish to incorporate, and thus deserves to be investigated. 

In order to be more even more explicit regarding the \textit{simplicity} of our approach to develop constitutive theory for binary mixtures, we wish to mention that there are other thermodynamic approaches used to develop models for binary fluid mixtures involving the dissipation due to drag force between the individual fluids. For example, using the principles of rational thermodynamics, Rajagopal and Tao~\cite{rajagopal.kr.tao.l:mechanics} (see also an earlier study \cite{alsharif1993}) obtained the following constitutive equations for the individual Cauchy stresses $\TT_1$ and $\TT_2$ (expressed in terms of partial pressures $p_1$, $p_2$ and the velocities $\vv_1$, $\vv_2$ of individual fluids): 
\begin{align}
\nonumber
        \TT_1 & = \left( - p_1 + \mathrm{c}_1 \div\vv_1 + \mathrm{c}_2 \div\vv_2\right)\II + 2\mathrm{c}_3 \DD(\vv_1) + 2\mathrm{c}_4 \DD(\vv_2) + \mathrm{c}_5 \VV_{12}, \\
        \nonumber
        \TT_2 & = \left( - p_2 + \mathrm{c}_6 \div\vv_1 + \mathrm{c}_7 \div\vv_2\right)\II + 2\mathrm{c}_8 \DD(\vv_1) + 2\mathrm{c}_9 \DD(\vv_2) - \mathrm{c}_5 \VV_{12}, 
\end{align}
where $\mathrm{c}_1,\dots,\mathrm{c}_9$ are the material coefficients and  $\VV_{12}:= \frac{\nabla \vv_1 - (\nabla \vv_1)^T}{2} - \frac{\nabla \vv_2 - (\nabla \vv_2)^T}{2}$ denotes the relative spin; the other symbols are introduced below in Section \ref{sec-basics}. Thus, the constitutive equations for $\TT_1$ and $\TT_2$ are of complicated forms (despite the fact that they depend on the velocities $\vv_1$ and $\vv_2$ linearly) and include $9$ material coefficients that is difficult to specify/experimentally measure. The intention of the study by M\'{a}lek and Rajagopal~\cite{malek-2008} and also of this paper is to provide the models with minimal number of model parameters that are experimentally measurable. We refer to the former reference~\cite{malek-2008} for a more detailed discussion regarding this issue. 


Besides M\'{a}lek and Rajagopal~\cite{malek-2008}, this study is also closely related to Sou\v{c}ek et al.~\cite{soucek-et-al-2014} There the authors studied (chemically non-reacting) binary mixtures focusing however on the comparison of the resulting detailed model with the balance equations for a single continuum. In doing so, the complete description for the binary mixture was given in terms of the mixture density, the barycentric velocity, the concentration of one component, the diffusive flux and the whole-mixture free energy. The constitutive equation for the Helmholtz free energy, in contrast with this study, contains an additional kinetic energy term associated with the diffusive mass flux. As a consequence, the model developed in Sou\v{c}ek et al.~\cite{soucek-et-al-2014}, in addition to the standard closure relations, leads to an evolution equation for the diffusive mass flux. Here, we stick more to the primary quantities such as the densities, the velocities, the energies and the entropies of the individual constituents and, consequently, we do not involve the diffusive kinetic energy in the fundamental thermodynamic relation (constitutive equation for the Helmholtz free energy). In this regard, this study thus provides an alternative view-point concerning the development of models for binary mixtures. 
In addition to that, our final intention is to also derive models in which all admissible flows associated with the whole mixture are isochoric. It means that the velocity of the mixture (given by averaging of the velocities of the individual constituents  by mass, molar, or volume fractions) has zero divergence.

Last, but not least, let us mention the notorious challenge to all theories of multi-component materials consisting in an identification of appropriate boundary conditions. This issue has been recognized as perhaps the biggest obstacle in solving real-world problems by means of mixture theory\cite{Rajagopal-and-Tao-1995}. In this regard, the model developed here provides a straightforward identification of the boundary conditions for the individual constituents (fluids) {\it from the boundary conditions associated with the whole-mixture velocities and stresses}, usually used for a single-component continuum. This, for example, allows one to equip the model with the traditional generalized slip condition (covering no-slip, full slip and Navier slip). This surprising property  follows from the fact that, in the approach developed in this study, the Cauchy stresses associated with the individual constituents can be directly related to the whole-mixture Cauchy stress. 

The structure of the paper is the following. In the next section, we briefly summarize the basic balance equations of the theory of interacting continua consisting of $N$ constituents. We also introduce mass, molar and volume fractions and associated notions of whole-mixture velocity weighted by these quantities. Starting from Section \ref{sec-binary-mixture}, we restrict ourselves to binary fluid mixtures, i.e. we set $N{=}2$, and we focus on the reformulation of those governing equations that will be later needed when applying the thermodynamical approach. Then, in Section \ref{sec:4}, we formulate the constitutive equations for the Helmholtz free energy and study the consequences of this assumption regarding the admissible form of the rate of entropy production. In Section \ref{sec-nondisspative},  we derive the model that is obtained as a consequence of the assumption that the mixture as a whole does not produce any entropy. In doing so, we obtain a complete description of a particular binary mixture model, which can be viewed as an counterpart of the Euler system in the case of a single compressible fluid. In Section \ref{sec-constitutive}, we present a thermodynamic framework for identification of the constitutive relations for a model (a closed system of governing partial differential equations) for binary fluid mixtures assuming that the entropy production corresponding to that of a single-component heat-conducting fluid (with dissipative mechanisms formulated in terms of the whole-mixture velocity) and with one additional mechanical term arising from internal friction due to mutual interaction of individual fluids. For lucidity, we summarize the complete closed system of governing equations that followed from this approach in Section \ref{sec:7}. 
Then, in Section \ref{sec-partical-cauchy-equilibrium}, we show that the choice of molar fraction as the weighting function in the definition of the whole-mixture velocity is compatible with the model of mixture of ideal gases in thermodynamic equilibrium. In Section \ref{sec-incompressible}, we repeat the whole derivation under the additional assumption (constraint) that the whole mixture responses as an incompressible fluid. Section \ref{sec:9} is devoted to the derivation of the boundary condition involving the velocity and Cauchy stress of the individual fluid from the knowledge of the boundary conditions for the whole-mixture velocity and whole-mixture Cauchy stress. The article ends with concluding remarks given in Section \ref{sec-summary}, list of references and Appendix, briefly recalling the description of mixtures of ideal gases.

\section{Basics of the theory of interacting continua}
\label{sec-basics}
This section recalls the basic governing equations of  the theory of interacting continua suitable to describe mechanical, chemical and thermal processes in mixtures of $N$ interacting fluids. We also introduce the mass, molar and volume fractions and for each of them we define the associated weighted whole-mixture velocity.  

\subsection{Balance equations}
The cornerstone of the theory of interacting continua is the assumption of co-occupancy (co-existence) stating that all constituents are present (co-exists) at all points in the current state of the body. Based on this assumption, the balance equations for mass, linear and angular momenta, energy and  entropy are formulated in the form of systems of partial differential equations (PDEs) that, besides the presence of interaction terms is the same as in the case of single continuum. Various levels of complexity are possible in mixture theories in terms of the employed level of detail involved in formulation of the balance equations, (see e.g. the classification by Hutter and J\"ohnk \cite{hutter.k.johnk.k:continuum}). Here, we require that the balances equations for mass, linear and angular momenta hold for each constituent of the mixture, while the balance equations for energy and entropy are considered in a summarized form for the whole mixture. More specifically, in our setting, the basic governing equations are those expressing:
\begin{itemize}
\begin{subequations}
\item Balance of mass for the individual constituent (labeled by $\alpha$) 
\begin{align}
\label{eq:framework-mass}
	\frac{\pa\rho_\alpha}{\pa t} + \div(\rho_\alpha\vv_\alpha) &= 
m_\alpha\,,\hspace{1cm}\alpha=1,\dots,N\,,
\end{align}
where $\rho_\alpha$ and $\vv_\alpha$ denote respectively the density and the velocity associated with the $\alpha$ constituent and $m_\alpha$ denotes the mass production (gain/loss) of the $\alpha$ component due to (chemical) reactions with the remaining constituents.
\item  Balance of linear momentum for the $\alpha$ constituent 
\begin{align}
	\label{eq:framework-momentum}
	\frac{\pa(\rho_\alpha\vv_\alpha)}{\pa t} + \div(\rho_\alpha\vv_\alpha\otimes\vv_\alpha) &= \div\TT_\alpha + \rho_\alpha\bb_\alpha + \vecI_\alpha + m_\alpha\vv_\alpha\,,\hspace{1cm}\alpha=1,\dots,N\,,
\end{align}
where $\TT_\alpha$, $\bb_\alpha$ and $\vecI_\alpha$ denote respectively the Cauchy stress, external body force and the interaction force of the $\alpha$ component.
\item  Balance of angular momentum for the $\alpha$ constituent \\ is reduced to the the statement concerning the symmetry of the Cauchy stress $\TT_\alpha$ for each constituent, i.e., 
\begin{align}
	\label{eq:framework-angular-momentum}
	\TT_\alpha = \TT_\alpha^\mathrm{T}\,,\hspace{1cm}\alpha=1,\dots,N\,.
\end{align}
This means that we consider a mixture consisting of non-polar constituent. \\
The above relations \eqref{eq:framework-mass} and \eqref{eq:framework-momentum} are supplemented with the constraints on the interaction terms - they must vanish, when summed over all constituents:
\begin{align}
\label{eq:framework-constraints}
	\suma m_\alpha = 0\,,\hspace{1cm}\suma (\vecI_\alpha + m_\alpha\vv_\alpha) = {\bf 0}\,.
\end{align}
\noindent
The first equation states that the total mass of the whole mixture is conserved although the mass of individual constituent can vary due to chemical reactions. The second equation represents the action-reaction principle of classical mechanics.
\item Balance of (total) energy for the whole mixture formulated as the sum of balance equations for the (total) energy of individual constituents
\begin{align}
	\label{eq:framework-total-energy}
	\frac{\pa}{\pa t}\left( \suma\rho_\alpha E_\alpha \right) + \div\left(\suma\rho_\alpha E_\alpha \vv_\alpha\right) &= \div\suma(\TT_\alpha\vv_\alpha - \qq_\alpha) + \suma\rho_\alpha\bb_\alpha\cdot\vv_\alpha + \suma\rho_\alpha r_\alpha\,.
\end{align}
Here $E_\alpha = e_\alpha + \frac{1}{2}|\vv_\alpha|^2$, $\qq_\alpha$ and $r_\alpha$ are respectively the total energy, 
the (non-convective) energy flux and the outer energy supply of the $\alpha$ constituent. \\
The thermodynamic approach presented below is based on a different form of the balance of energy, which we obtain from the (total) energy balance \eqref{eq:framework-total-energy} by subtracting the balance of kinetic energy. The latter is obtained by multiplying (\ref{eq:framework-momentum}) by $\vv_\alpha$ followed by  straightforward manipulations (using also \eqref{eq:framework-mass} and \eqref{eq:framework-angular-momentum}) resulting at:
\be
\label{eq:framework-kinetic-energy}
	\frac{\pa}{\pa t}\left(\frac{\rho_\alpha|\vv_\alpha|^2}{2}\right) + \div\left(\frac{\rho_\alpha|\vv_\alpha|^2}{2}\vv_\alpha\right) = \div\left(\TT_\alpha\vv_\alpha\right) -\TT_\alpha:\DD(\vv_\alpha) + \rho_\alpha\bb_\alpha\cdot\vv_\alpha + \vecI_\alpha\cdot\vv_\alpha + m_\alpha\frac{|\vv_\alpha|^2}{2}\,,
\ee
where $\DD(\aa)$ denotes the symmetric part of the gradient of a vector $\aa$, i.e., 
\begin{align*}
\DD(\aa) \eqdef \frac{1}{2}(\nabla\aa+{(\nabla\aa)}^\mathrm{T})\,. 
\end{align*}
Summing \eqref{eq:framework-kinetic-energy} over $\alpha$, $\alpha=1, \dots, N$, and subtracting the result from (\ref{eq:framework-total-energy}), we get, as a consequence of the balance of energy, the balance equation for the internal energy in the form 
\begin{align}
\label{eq-internal-energy-balance}
	\frac{\pa}{\pa t}\left(\suma\rho_\alpha e_\alpha\right) + \div\left(\suma\rho_\alpha e_\alpha\vv_\alpha\right) &= -\div\left(\suma\qq_\alpha\right) + \suma\rho_\alpha r_\alpha + \suma\TT_\alpha{:}\DD(\vv_\alpha)
	 - \suma\left( \vecI_\alpha\cdot\vv_\alpha +  m_\alpha\frac{|\vv_\alpha|^2}{2}\right)\,.
\end{align}
\item Balance of entropy for the whole mixture again postulated as the sum of balance equations for the entropy for  individual constituents:
\begin{align}
\label{eq-entropy-balance}
\frac{\pa}{\pa t}\left( \suma \rho_\alpha\eta_\alpha\right) + \div\left(\suma\rho_\alpha\eta_\alpha\vv_\alpha\right) = -\div\left(\suma\entropyflux_\alpha\right) + \suma \rho_\alpha\entropysupply_\alpha + \entropyproduction\,,
\end{align}
where $\eta_\alpha$, $\entropyflux_\alpha$, $\entropysupply_\alpha$ denote respectively the entropy, the entropy flux and the outer entropy supply of the $\alpha$ constituent, and $\entropyproduction\eqdef\suma\entropyproduction_\alpha$ is the total entropy production for the whole mixture - a sum of the individual entropy productions of all the constituents. Requiring that 
\begin{align}
\label{eq-second-law}
\entropyproduction\geq 0\,,
\end{align}
\end{subequations}
the validity of the second law of thermodynamics for the whole mixture is fulfilled.
\end{itemize}

\subsection{Mass,  molar and volume fractions and the whole mixture velocities}
\label{sec-definitions}
We first define the {\it whole-mixture density} $\rho$ and the associated {\it mass fractions (concentrations)} $c_\alpha$
through
\begin{align}
\label{def-rho-c-alpha}
\rho\eqdef\suma\rho_\alpha\,,\hspace{1cm}c_\alpha\eqdef\frac{\rho_\alpha}{\rho}\,,\hspace{1cm}\alpha=1,\dots,N\,.
\end{align}
If the $\alpha$ constituent has a {\it molar mass} $M_\alpha$, then the corresponding {\it molar concentrations} $\cm_\alpha$ and the {\it whole-mixture molar concentration} $\cm$ are given by 
\begin{align}
\label{def-cM-alpha-cM}
\cm_\alpha\eqdef \frac{\rho_\alpha}{M_\alpha}\,,\hspace{1cm}\alpha=1,\dots,N\,,\hspace{1cm}\cm\eqdef\suma\cm_\alpha\,,
\end{align}
and the {\it molar fractions} $\mf_\alpha$ by
\begin{align}
\mf_\alpha\eqdef\frac{\cm_\alpha}{\cm}\,,\hspace{1cm}\alpha=1,\dots,N\,.
\end{align}
We assume that all the constituents of the mixture are properly accounted for in the above definitions of the mixture density \eqref{def-rho-c-alpha} and the mixture molar concentration \eqref{def-cM-alpha-cM}, i.e. we are requiring the validity of the {\it mass additivity constraint} and {\it molar additivity constraint}. As a direct consequence of the above definitions we then obtain that
\begin{equation}
\label{def-mass-molar-additivity}
\suma c_\alpha = 1\qquad \textrm{ and } \qquad\suma x_\alpha = 1\,.
\end{equation}
In many physically relevant situations (e.g. if the constituents do not mix at the molecular level, as for instance, in the case of the emulsions) it makes sense to introduce {\it volume fractions} $\phi_\alpha$ by
\begin{equation}
\phi_\alpha \eqdef \frac{\rho_\alpha}{\rhot_\alpha}\,,\hspace{1cm}\alpha=1,\dots,N\,,
    \end{equation}
where $\rhot_\alpha$ denotes the true {\it material density} of the $\alpha$ constituent, i.e. the density one would measure for a pure $\alpha$ substance. Under the assumption of saturated mixture without any voids, a counterpart of the relations \eqref{def-mass-molar-additivity}, called {\it volume additivity constraint} reads as
\begin{equation}
\label{def-volume-additivity}
\suma \phi_\alpha = 1\,.
\end{equation}
Under the assumptions above the volume fraction $\phi_\alpha$ indeed expresses the partial volume occupied locally by the $\alpha$ constituent, while the volume additivity constraint \eqref{def-volume-additivity} expresses the molecular non-mixing and saturation (i.e. absence of voids). 

Let us now consider a generic {\it weight functions} $\weight_\alpha$, $\alpha=1,\dots,N$, such that
\begin{align}
\label{eq-suma-weight}
\suma\weight_\alpha = 1\,,
\end{align}
and define the whole-mixture velocity as a corresponding $\weight_\alpha$-weighted average of the  velocities of the individual constituents, i.e.,
\begin{equation}
\label{def-vmix}
\vvmix \eqdef \suma\weight_\alpha\vv_\alpha\,.
\end{equation}
We observe that the three special choices of the weights, namely $\weight_\alpha{=}c_\alpha$, $\weight_\alpha{=}x_\alpha$ and $\weight_\alpha{=}\phi_\alpha$, lead to the following quantities:
\begin{equation}
\label{def-velocities}
\vv\eqdef \suma c_\alpha\vv_\alpha\,,\qquad \qquad \vvm\eqdef\suma\mf_\alpha\vv_\alpha\qquad \textrm{ and } \qquad
{\vvphi} \eqdef \suma\phi_\alpha\vv_\alpha\,.
\end{equation}
The quantity $\vv$ is the barycentric velocity of the mixture, used the most traditionally as a definition of the mixture velocity. The other two definitions of the whole-mixture velocity are perhaps less popular, albeit the volume averaged $\vvphi$ has been employed in the mathematical treatments of quasi-incompressible materials, see \cite{abels2012,abels2013}.

To a general definition of the whole-mixture velocity $\vvmix$, we assign an associated {\it diffusive mass flux} $\jjmix_\alpha$ by
\begin{equation}
\label{def-diff-fluxes}
\jjmix_\alpha \eqdef \rho_\alpha(\vv_\alpha - \vvmix)\,,\hspace{1cm}\alpha=1,\dots,N\,.
\end{equation}
Here and in what follows the superscript $^{\textrm{mix}}$ is added to the quantities that depend on the weights $\weight_\alpha$, and consequently they are different if we consider mass, molar or volume fractions.

\section{Governing balance equations for binary fluid mixtures}
\label{sec-binary-mixture}

In the remaining parts of this study we consider binary fluid mixtures, i.e. we restrict ourselves to the case when $N{=}2$. In this section, we rewrite the governing equations in this simplified setting focusing on the balance equations for mass of individual constituents, and for the balance equation for the whole-mixture internal energy and for the whole-mixture entropy, as these are the equations that enter into the thermodynamical approach presented in Section \ref{sec:4}. The derivation is performed for a general whole-mixture velocity $\vvmix$, introduced in \eqref{def-vmix}.

Thus, as $N{=}2$, setting
\begin{align}
\label{def-relabel0}
\weight\eqdef \weight_1\,, \qquad 
m\eqdef m_1\qquad\textrm{ and } \qquad \vecI\eqdef\vecI_1\,,
\end{align}
it follows from \eqref{eq-suma-weight}  and \eqref{eq:framework-constraints} that 
\begin{align}
\label{def-relabel1}
\weight_2 = 1-\weight\,, \qquad
m_2 =  - m  \qquad \textrm{ and } \qquad \vecI_2 = - \vecI - m (\vv_1 - \vv_2)\,.
\end{align}
Then the last sum in \eqref{eq-internal-energy-balance} simplifies to 
\begin{align}
\label{def-relabel2}
\sumad\left( \vecI_\alpha\cdot\vv_\alpha +  m_\alpha\frac{|\vv_\alpha|^2}{2}\right) = \left(\vecI + \frac{m}{2}(\vv_1 - \vv_2)\right)\cdot (\vv_1 - \vv_2)\,.
\end{align}

Introducing further the notation
\be
\label{def-q-r-e}\begin{split}
	\qqi\eqdef\sumad\qq_\alpha\,,\qquad  
	\ei &\eqdef \frac{1}{\rho} \sumad \rho_\alpha e_\alpha\,, \qquad 
	\eta\eqdef \frac{1}{\rho} \sumad \rho_\alpha \eta_\alpha\,, \qquad
	\ri\eqdef \frac{1}{\rho}\sumad \rho_\alpha r_\alpha\quad\textrm{ and }\quad \entropysupply \eqdef \frac{1}{\rho}\sumad \rho_\alpha\entropysupply_\alpha\,,
	\end{split}
\ee
the balance equations \eqref{eq:framework-mass}, \eqref{eq-internal-energy-balance} and \eqref{eq-entropy-balance} (for binary mixtures) take the form\footnote{We do not list the balance equations of linear momentum for individual constituents here as these equations do not explicitly enter into the thermodynamic approach presented below. This is due to the fact that these equations have been used in Section 2 for the derivation of an alternative form of the balance of energy, namely the equation \eqref{eq-internal-energy-balance}.}
\begin{subequations}
\label{eqs-auxiliary-for-of-balances}
\begin{align}
\frac{\pa \rho_\alpha}{\pa t} + \div(\rho_\alpha\vvmix) &= (-1)^{\alpha+1} m - \div\jjmix_\alpha \qquad(\alpha=1,2) \qquad \textrm{ with } \begin{cases} \jjmix_1 = \rho_1 (1-\omega)(\vv_1-\vv_2), \\ \jjmix_2 = - \rho_2 \omega (\vv_1-\vv_2), \end{cases}\label{eq-aux00}\\
\frac{\pa}{\pa t}(\rho\ei) + \div\left(\rho\ei\vvmix\right) &= -\div\left(\sumad e_\alpha\jjmix_\alpha\right)-\div\qqi +\rho\ri + \sumad \TT_\alpha:\DD(\vv_\alpha) - \left(\vecI + \frac{m}{2}(\vv_1{-}\vv_2)\right)\cdot(\vv_1{-}\vv_2)\,, \label{eq-aux01} \\
\label{eq-entropy-balance-total-mass}
\frac{\pa(\rho\eta)}{\pa t} + \div\left(\rho\eta\vvmix\right) &= -\div\entropyfluxmix+ \rho\entropysupply + \entropyproduction\,,
\end{align}
\end{subequations}
where the whole-mixture entropy flux $\entropyfluxmix$ denotes
\begin{align}
\label{def-entropyflux-total-mass}
\entropyfluxmix\eqdef\sumad\left(\entropyflux_\alpha + \rho_\alpha \eta_\alpha(\vv_\alpha - \vvmix ) \right)\,.
\end{align}

Furthermore, we observe that the term $\sumad e_\alpha\jjmix_\alpha$ appearing in \eqref{eq-aux01} takes the form
\begin{align}\label{aux03}
\sumad e_\alpha\jjmix_\alpha = \EEEmix(\vv_1-\vv_2)\,,\hspace{1cm}\text{where}\hspace{1cm}\EEEmix\eqdef (\rho_1 e_1 (1-\weight) - \rho_2 e_2\weight)\,.
\end{align}
We split its divergence into an affine combination, i.e., for a scalar function $\gamma$ we write 
\begin{align}\label{aux04}
\div \left( \EEEmix(\vv_1-\vv_2)\right) = \gamma \EEEmix \div\vv_1 - \gamma \EEEmix \div\vv_2 + \nabla (\gamma\EEEmix) \cdot (\vv_1 - \vv_2) + \div \left( (1-\gamma) \EEEmix(\vv_1-\vv_2)\right)\,. 
\end{align}

Defining the partial mean normal stresses $\mns_\alpha$ and the deviatoric part $\mathbb{A}^d$ of a second-order tensor $\mathbb{A}$ through
\begin{align}
    \mns_\alpha \eqdef \frac13 \tr \TT_\alpha \quad \textrm{ and } \quad \mathbb{A}_\alpha^d \eqdef \mathbb{A} - \frac13 (\tr \mathbb{A}) \mathbb{I}, \qquad (\mathbb{I} \textrm{ is the identity tensor})
\end{align}
and inserting the above notation as well as the splitting \eqref{aux04} into \eqref{eq-aux01} we get 
\begin{equation}
\label{eq-internal-e-barycentric}
\begin{split}
  \frac{\pa}{\pa t}(\rho\ei) + \div\left(\rho\ei\vvmix\right) &= -\div \left( \qqi + (1-\gamma) \EEEmix(\vv_1-\vv_2)\right) +\rho\ri + (\mns_1{-}\gamma\EEEmix)\div\vv_1 + (\mns_2{+}\gamma\EEEmix)\div\vv_2 \\ & + \TT^d_1:\DD^d(\vv_1) + \TT^d_2:\DD^d(\vv_2)- \left(\vecI + \frac{m}{2}(\vv_1{-}\vv_2) + \nabla(\gamma\EEEmix)\right)\cdot(\vv_1{-}\vv_2)\,.
\end{split}
\end{equation}

\section{Specification of the energy storage mechanism and its consequences} \label{sec:4}
In this section, we postulate the constitutive equations for the whole-mixture Helmholtz free energy depending on the whole-mixture temperature field and the (partial) densities of the individual fluids. Then we inspect the consequences that this assumption implies regarding the admissible form for the rate of entropy production.

\subsection{Constitutive equation for the whole-mixture Helmholtz free energy}
\label{sec-fundamental-relation}
For each constituent, similarly as in the context of thermodynamics of single continuum, one can use $\eta_\alpha$ and $e_\alpha$ to introduce thermodynamic temperatures $\vartheta_\alpha$ associated with the individual constituents by $\vartheta_\alpha=\frac{\pa e_\alpha}{\pa \eta_\alpha}$. Then one can define the partial Helmholtz free energy $\psi_\alpha$ as the corresponding Legendre transform so that
\begin{equation}
\label{def-psi-alpha}
\psi_\alpha =  e_\alpha - \vartheta_\alpha \eta_\alpha\,\qquad\text{and}\qquad \frac{\pa\psi_\alpha}{\pa \vartheta_\alpha} = -\eta_\alpha\,.
\end{equation}
In this study, we shall assume that the (two) temperature fields coincide, so that
\begin{align}
\vartheta\eqdef \vartheta_1 = \vartheta_2\,.
\end{align}
Furthermore, we set
\begin{equation}
\label{def-psi-mix}
\psi\eqdef \frac{1}{\rho}\sumad\rho_\alpha\psi_\alpha\,,
\end{equation}
and, consequently, in accordance with the notation introduced in Section~\ref{sec-binary-mixture}, we get
\begin{equation}
\rho\psi = \rho e -  \vartheta \rho\eta\,.
\end{equation}
The fundamental thermodynamic relation, characterizing the energy storage mechanism, is here expressed in the form of a constitutive equation for $\rho\psi$ (volumetric Helmholtz free energy), being of the form
\be
\label{def-psi-compressible}
	\rho\psi = \widehat{\rho\psi}(\vartheta,\rho_1,\rho_2)\,,
\ee 
and in view of \eqref{def-psi-mix}, \eqref{def-psi-alpha}, and \eqref{def-q-r-e}, the standard relation between the total entropy and Helmholtz free energy holds
\begin{align}
\label{eq-aux3}
\rho\eta = -\frac{\pa\widehat{\rho\psi}}{\pa\vartheta}\,.
\end{align}
We define the {\it chemical potentials } $\mu_\alpha$ by
\begin{align}
\label{def-chem-pot}
\mu_\alpha \eqdef\frac{\pa\widehat{\rho\psi}}{\pa\rho_\alpha}\,,\quad\alpha=1,2\,,
\end{align}
and the {\it thermodynamic pressure} $p$ is introduced through the Euler relation, known from classical equilibrium thermodynamics \cite{callen.hb:thermodynamics} by
\begin{align}
\label{def-pressure}
p \eqdef -\rho\ei + \vartheta\rho\eta + \sumad\rho_\alpha\mu_\alpha\,.
\end{align}

\subsection{Consequences of the choice $\rho\psi=\widehat{\rho\psi}(\vartheta,\rho_1,\rho_2)$}
\label{sec-consequences}
Defining for any quantity $z$ the  {\it material time derivative} $\dot{\overline{z}}$ associated with the whole-mixture velocity $\vvmix$, through
\begin{align}
\label{def-dotz}
  \dot{\overline{z}} \eqdef \frac{\partial z}{\partial t} + \vvmix \cdot \nabla z\,,
\end{align}
the balances equations \eqref{eq-aux00}, \eqref{eq-internal-e-barycentric} and \eqref{eq-entropy-balance-total-mass} can be rewritten as follows
\begin{subequations}
\label{eqs-auxiliary-for-of-balances2}
\begin{align}
\dot{\overline{\rho_\alpha}} &= -\rho_\alpha\div\vvmix + (-1)^{\alpha+1} m - \div\jjmix_\alpha \qquad(\alpha=1,2)\,\qquad\textrm{ with } \begin{cases} \jjmix_1 = \rho_1 (1-\omega)(\vv_1-\vv_2)\,, \\ \jjmix_2 = - \rho_2 \omega (\vv_1-\vv_2)\,, \end{cases} \label{pepa:9995}\\
\dot{\overline{\rho e}} &= - \rho e \div\vvmix -\div \left( \qqi + (1-\gamma) \EEEmix(\vv_1-\vv_2)\right) +\rho\ri + (\mns_1{-}\gamma\EEEmix)\div\vv_1 + (\mns_2{+}\gamma\EEEmix)\div\vv_2 \nonumber\\ & + \TT^d_1:\DD^d(\vv_1) + \TT^d_2:\DD^d(\vv_2)- \left(\vecI + \frac{m}{2}(\vv_1{-}\vv_2) + \nabla(\gamma\EEEmix)\right)\cdot(\vv_1{-}\vv_2)\,,\\
\dot{\overline{\rho\eta}} &= -\rho\eta\div\vvmix -\div\entropyfluxmix+ \rho\entropysupply + \entropyproduction\,.
\end{align}
\end{subequations}
Next, applying the material time derivative to 
\begin{equation}
(\rho e)-\vartheta(\rho\eta) = \rho\psi = \widehat{\rho\psi}(\vartheta,\rho_1,\rho_2)\,,
\end{equation}
and using \eqref{eq-aux3}, \eqref{def-chem-pot}, and \eqref{def-pressure}, we get
\begin{align}
\label{eq-aux-12}
\dot{\overline{\rho\ei}} = \vartheta\dot{\overline{\rho\eta}} + \sumad\mu_\alpha\dot{\overline{\rho_\alpha}}\,.
\end{align} 
As a next step, we intend to substitute the time derivatives from \eqref{eqs-auxiliary-for-of-balances2} into the last identity. This will in particular lead to the term $\sumad \mu_\alpha \div\jjmix_\alpha$ that we rewrite as
\begin{equation}
\label{aux-60}
\sumad \mu_\alpha \div\jjmix_\alpha = \div\left(\sumad\mu_\alpha\jjmix_\alpha\right) - \sumad\jjmix_\alpha\cdot\nabla\mu_\alpha\,.
\end{equation}
Introducing new symbols $\mu$, $\mumix$ and $\mumixvec$ through
\begin{align}
\label{def-mumixvec}
\mu\eqdef\mu_1 - \mu_2\,,\qquad\qquad
 \mumix\eqdef \rho_1(1{-}\weight)\mu_1 - \rho_2\weight\mu_2\,,\qquad\qquad
 \mumixvec\eqdef \rho_1(1{-}\weight)\nabla\mu_1 - \rho_2\weight\nabla\mu_2\,, 
\end{align}
we get
\begin{align}
\label{aux-61}
    \sumad\mu_\alpha \jjmix_\alpha = \mumix(\vv_1-\vv_2)\,\qquad\text{and}\qquad
    \sumad\jjmix_\alpha\cdot\nabla\mu_\alpha =\mumixvec\cdot(\vv_1-\vv_2)\,.
\end{align}
Thus, upon inserting \eqref{eqs-auxiliary-for-of-balances2} into \eqref{eq-aux-12}, 
using \eqref{aux-60}, \eqref{aux-61} and the formula
\begin{align}
\nonumber
\div\vvmix &= \div(\weight\vv_1+(1{-}\weight)\vv_2) = \weight\div\vv_1{+}(1{-}\weight)\div\vv_2 + (\vv_1{-}\vv_2)\cdot\nabla \weight\,,
\end{align}
we conclude, after some manipulations, that
\begin{align}
\nonumber
	& 
    -\div\entropyfluxmix + 	\rho\entropysupply + \entropyproduction =
	-\div\left(\frac{\qqi+\left((1-\gamma)\EEEmix -\mumix\right)(\vv_1-\vv_2)}{\vartheta}\right)+\frac{\rho\ri}{\vartheta}\\\nonumber &+ \frac{1}{\vartheta}\Bigg\{  (\mns_1{-}\gamma\EEEmix{+}\weight p)\div\vv_1 + (\mns_2{+}\gamma\EEEmix{+}(1{-}\weight)p)\div\vv_2 + \TT^d_1:\DD^d(\vv_1) + \TT^d_2:\DD^d(\vv_2)  -m \mu  \\\label{eq-entropy-bal-final} &  -\left(\frac{\qqi+\left((1-\gamma)\EEEmix -\mumix\right)(\vv_1-\vv_2)}{\vartheta}\right)\cdot\nabla\vartheta - \left(\vecI + \frac{m}{2}(\vv_1{-}\vv_2) + \nabla(\gamma\EEEmix) - p\nabla \weight + \mumixvec  \right)\cdot(\vv_1{-}\vv_2)
	\Bigg\}\,.
\end{align}
Comparing the left-hand side and the right-hand side of this identity, leads us to identify the entropy flux, entropy supply and the entropy production in the following way
\begin{subequations} 
\label{pepa:999}
\begin{align}
\entropyfluxmix = \frac{\qqi+((1{-}\gamma)\EEEmix - \mumix)(\vv_1-\vv_2)}{\vartheta} \qquad\text{ and }\qquad
\entropysupply = \frac{\ri}{\vartheta}\,,
\end{align}
and
\begin{align}
\nonumber
\xi\eqdef \vartheta\entropyproduction &=   (\mns_1{-}\gamma\EEEmix{+}\weight p)\div\vv_1 + (\mns_2{+}\gamma\EEEmix{+}(1{-}\weight)p)\div\vv_2 + \TT^d_1:\DD^d(\vv_1) + \TT^d_2:\DD^d(\vv_2)  -m \mu  \\\label{eq-entropy-prod-final} &  -\left(\frac{\qqi+\left((1-\gamma)\EEEmix -\mumix\right)(\vv_1-\vv_2)}{\vartheta}\right)\cdot\nabla\vartheta - \left(\vecI + \frac{m}{2}(\vv_1{-}\vv_2) + \nabla(\gamma\EEEmix) - p\nabla \weight + \mumixvec  \right)\cdot(\vv_1{-}\vv_2)\,.
\end{align}
\end{subequations}
\section{Special case - binary fluid mixture with zero entropy production}
\label{sec-nondisspative}

Once we postulated how the material stores the energy (see \eqref{def-psi-compressible}), we should provide information how the material produces the entropy. We first look at the consequences regarding the form of governing equations in the case when the mixture as a whole produces \textit{no} entropy, i.e. $\xi = 0$. It then follows from \eqref{pepa:999} that this happens if the quantities $m$, $\TT_1$, $\TT_2$, $\vecI$ and $\qqi$
satisfy
\begin{equation}
    \begin{split}
    m&=0\,,\qquad \mns_1 = {-}\weight p + \gamma\EEEmix\,, \quad \mns_2 = - (1{-}\weight)p - \gamma\EEEmix\,, \qquad \TT^d_1 = {\mathbf{0}}\,, \quad \TT^d_2 = \mathbf{0}\,,\\ 
    \vecI &= - \nabla(\gamma\EEEmix) + p\nabla \weight - \mumixvec\,, \qquad \qqi = - \left((1-\gamma)\EEEmix -\mumix\right)(\vv_1{-}\vv_2)\quad\implies\quad \entropyfluxmix = \mathbf{0}\,.
    \end{split} \label{pepa:998}
\end{equation}
Note that due to \eqref{def-psi-compressible} it follows from \eqref{eq-aux3}, \eqref{def-chem-pot} and \eqref{def-pressure} that 
\begin{equation}
    p = p (\vartheta, \rho_1, \rho_2)\,, \qquad \mumix = \mumix(\vartheta, \rho_1, \rho_2)\,, \qquad \mumixvec = \mumixvec(\vartheta, \rho_1, \rho_2)\,.\label{pepa:997}
\end{equation}
By looking at the particular choices of $\vvmix$ given in \eqref{def-velocities}, we also observe that 
\begin{equation}
    \vvmix= \vvmix(\vartheta, \rho_1, \rho_2,\vv_1,\vv_2)\,.\label{pepa:996}
\end{equation}
Note that the dependence of $\vvmix$ on $\vartheta$ appears if the weights are the volume fractions and the material (true) densities are non-constants, but are given by additional state equations of the form $\rhot_\alpha = \widehat{\rhot_\alpha}(\vartheta)$.

Upon inserting \eqref{pepa:998} into \eqref{eq:framework-mass}, \eqref{eq:framework-momentum} and also \eqref{eqs-auxiliary-for-of-balances2} we obtain the closed system of governing equations for the unknowns $\rho_1$, $\rho_2$, $\vv_1$, $\vv_2$ and $e$ (or $\vartheta$): 
\begin{subequations}
\label{ideal-bin-mixture}
\begin{align}
\frac{\pa \rho_\alpha}{\pa t} + \div(\rho_\alpha\vv_\alpha) & =  0 \,\quad\qquad\qquad (\alpha=1,2)\,, \label{eq-mass-idealmix}\\
\frac{\pa(\rho_\alpha\vv_\alpha)}{\pa t} + \div(\rho_\alpha\vv_\alpha\otimes\vv_\alpha) &= \rho_\alpha\bb_\alpha + \mathbf{f}_{\alpha} \qquad  (\alpha=1,2) \qquad \textrm{ with } \begin{cases} \mathbf{f}_1 = - \weight \nabla p - \mumixvec\,, \\
\mathbf{f}_2 = - (1- \weight) \nabla p + \mumixvec\,,
\end{cases} \label{eq-momentum-idealmix}\\
\frac{\pa}{\pa t}(\rho\ei) + \div\left(\rho\ei\vvmix\right) &=  -p \div\vvmix + \rho\ri - \sumad\mu_\alpha\div(\rho_\alpha(\vv_\alpha{-}\vvmix))\,. \label{eq-energy-idelamix} 
\end{align}
\end{subequations}
(Recall that $r$ and $\bb_\alpha$, $\alpha = 1,2$, are given external sources (such as radiation or gravity).) In addition, the following equation, that is a consequence of \eqref{eq-entropy-balance}, can be added to \eqref{ideal-bin-mixture}
\begin{align} 
\label{eq-entropy-idealmix}
\frac{\pa(\rho\eta)}{\pa t} + \div\left(\rho\eta\vvmix\right) &=  \frac{\rho\ri}{\vartheta}\,.
\end{align}
Also, instead of \eqref{eq-mass-idealmix}, we could use \eqref{eq-aux00} and consider the following equations
\begin{align}
\frac{\pa \rho_\alpha}{\pa t} + \div(\rho_\alpha\vvmix) & =  - \div\jjmix_\alpha \qquad(\alpha=1,2) \qquad \textrm{ with } \begin{cases} \jjmix_1 = \rho_1 (1-\omega)(\vv_1-\vv_2), \\ \jjmix_2 = - \rho_2 \omega (\vv_1-\vv_2). \end{cases}\label{eq-mass-idealmix2}
\end{align}

\section{Specification of the entropy production and derivation of the constitutive relations}
\label{sec-constitutive}
In this section, we present a thermodynamic framework for identification of the constitutive relations for a model (a closed system of governing partial differential equations) for binary fluid mixtures. 
To this aim, we adopt a minimalistic approach, describing the mixture almost entirely in terms of the quantities associated with the mixture as a whole. In particular, our minimalistic approach will lead to entropy production corresponding to that of a single-component heat-conducting fluid (with dissipative mechanisms formulated in terms of $\vvmix$ ), augmented by one additional mechanical term arising from internal friction due to mutual interaction of the two fluids. In order to identify the constitutive relations, we employ formula \eqref{eq-entropy-bal-final}, inferred from the governing balance equations and from the specification of the energy storage mechanism, as a constraint in a thermodynamical approach known as maximization of the rate of entropy production \cite{Rajagopal-Srinivasa-2004}. As observed in Remark \ref{remark-dissipation-potential} below, this last step can be equivalently replaced by formulating the irreversible part of material response in terms of a suitable dissipation potential.

\subsection{Specification of rate of entropy production}
\label{sec-entrprod-ansatz}
Let us now formulate the second constitutive assumption concerning the considered binary fluid mixture by specifying the way how the whole mixture produces the entropy. We choose a particularly simple form of the entropy production, which differs from that of a single-component compressible heat conducting fluid\cite{Malek2016} only by two terms - a chemical reaction term and a term due to dissipation by friction due to mutual motions of the two fluids. Thus, we assume that 
\be
\label{eq:diss}
	\hat{\xi} = \frac{3\lambda{+}2\nu}{3}(\div\vvmix)^2+2\nu |\DD^d(\vvmix) |^2 + \kappa|\nabla\vartheta|^2+\beta\mu^2 + \alpha|\vv_{1}{-}\vv_2|^2\,,
\ee 
where
\be
	3\lambda{+}2\nu{>}0,\quad\nu{>}0,\quad\kappa{>}0,\quad\beta{>}0,\quad\alpha{>}0\,,
\ee
which ensure non-negativity of the entropy production in accordance with the second law of thermodynamics.

Note that all material parameters can be functions of the weight $\weight$ (and of course other state variables). A particularly simple choice would be 
\begin{subequations}
\label{param-weighting}
\begin{align}
\lambda = \weight\lambda_1+(1-\weight)\lambda_2\,\qquad 
\nu = \weight\nu_1+(1-\weight)\nu_2\,\qquad 
\kappa = \weight\kappa_1+(1-\weight)\kappa_2\,,
\end{align}
where $\lambda_\alpha$, $\nu_\alpha$ and $\kappa_\alpha$, $\alpha=1,2$ are the bulk and viscosities and thermal conductivities of the pure constituents. In order to ensure that the interaction parameters in the absence of one of the constituents vanish, it is also natural to assume that
\begin{align}
\beta = \weight(1-\weight)\tilde{\beta}\,,\qquad\alpha = \weight(1-\weight)\tilde{\alpha}\,.
\end{align}
\end{subequations}
The above choice (but clearly not just this one), provides consistency of the final mixture model with the single-component model of pure constituents in the limit of one of the substances vanishing, see the Remark~\ref{remark-limiting-to-pure-states}.

Note that we consider a very simple piece-wise quadratic form even for the chemical reaction term ($\beta\mu^2$), which will consequently result into linear chemical kinetics. This is clearly an oversimplification which can be easily circumvented by a more suitable choice of the chemical-reaction contribution to the entropy production, see e.g. Bothe and  Dreyer\cite{bothe-2015}.

\subsection{Constitutive relations derived by maximization of the rate of entropy production}
\label{sec-maximization}
Let us now identify the constitutive relations by employing the principle of {\it maximization of rate of entropy production} formulated by Rajagopal and Srinivasa in \cite{Rajagopal-Srinivasa-2004}, who argued that \textit{``in entropy-producing processes, a
specific choice from among a competing class of constitutive functions can be made so that the state variables evolve in a way that maximises the rate of entropy production''}.
In our setting, we maximize the entropy production $\hat{\xi}$ given in \eqref{eq:diss} with respect to the ``two-component'' affinities $\div\vv_1$, $\div\vv_2$, $\DD^d(\vv_1)$, $\DD^d(\vv_2)$, $\nabla\vartheta$, $\mu$, $\vv_1{-}\vv_2$ subject to a constraint that the dissipation also equals $\xi$ given by \eqref{eq-entropy-prod-final}, i.e. we solve the following constrained optimization problem:
\begin{align}
\label{eq-maximization-mass}
	\underset{\text{w.r.t } \div\vv_1,\div\vv_2,\DD^d(\vv_1),\DD^d(\vv_2),\nabla\vartheta,\mu,\vv_1{-}\vv_2}{\text{maximize}}\left\{ \hat{\xi} + \ell(\hat{\xi}-\xi)\right\}\,,
	\end{align}
where $\ell$ denotes the Lagrange multiplier associated with the constraint \eqref{eq-entropy-prod-final}.

The definition of the mixture velocity $\vvmix$, see \eqref{def-vmix}, implies the following identities:
\begin{subequations}
\label{pepa:20}
\begin{align}
	\nabla\vvmix &= \weight\nabla\vv_1 + (1{-}\weight)\nabla\vv_2 + (\vv_1{-}\vv_2)\otimes\nabla \weight\,,\label{pepa:20a}\\
    \div\vvmix &= \weight\div\vv_1+(1{-}\weight)\div\vv_2 + (\vv_1{-}\vv_2)\cdot\nabla \weight\,,\label{pepa:20b}\\
    \DD^d(\vvmix) &= \weight\DD^d(\vv_1) + (1{-}\weight)\DD^d(\vv_2) + \frac{1}{2}\left((\vv_1{-}\vv_2)\otimes\nabla \weight + \nabla \weight\otimes(\vv_1{-}\vv_2)\right) - \frac{1}{3}(\vv_1{-}\vv_2)\cdot\nabla \weight\ \II\,.\label{pepa:20c}
    \end{align}
\end{subequations}
Using \eqref{pepa:20}, we directly conclude from \eqref{eq:diss} that
\begin{subequations}
\label{eq-xi-derivatives}
\begin{alignat}{3}
\frac{\pa\hat{\xi}}{\pa\div\vv_1} &= 2\weight\left(\frac{3\lambda+2\nu}{3}\right)\div\vvmix\,, \hspace{4.3cm}\frac{\pa\hat{\xi}}{\pa\div\vv_2}&&=2(1{-}\weight)\left(\frac{3\lambda+2\nu}{3}\right)\div\vvmix\,,\\
\frac{\pa\hat{\xi}}{\pa\DD^d(\vv_1)} &= 4\nu \weight\DD^d(\vvmix)\,, \hspace{5.5cm}
\frac{\pa\hat{\xi}}{\pa\DD^d(\vv_2)} &&= 4\nu(1{-}\weight)\DD^d(\vvmix)\,, \\
\frac{\pa\hat{\xi}}{\pa\nabla\vartheta} &= 2\kappa\nabla\vartheta\,, 
\hspace{7cm}\frac{\pa\hat{\xi}}{\pa\mu} &&= 2\beta\mu\,,\\
\frac{\pa\hat{\xi}}{\pa(\vv_1{-}\vv_2)} 
&=
4\nu\DD^d(\vvmix)\nabla\weight + 2\left(\frac{3\lambda{+}2\nu}{3}\right)\div\vvmix\nabla\weight + 2\alpha(\vv_1{-}\vv_2)\,.
\end{alignat}
\end{subequations}
Thus, the necessary conditions for the constrained maximization \eqref{eq-maximization-mass} take the form
\begin{subequations} \label{pepa:995}
\begin{alignat}{2}
 \frac{1{+}\ell}{\ell}2\weight\left(\frac{3\lambda{+}2\nu}{3}\right)\div\vvmix &= \mns_1{-}\gamma\EEEmix{+}\weight p\,,\hspace{1cm}
 	\frac{1{+}\ell}{\ell}2(1{-}\weight)\left(\frac{3\lambda{+}2\nu}{3}\right)\div\vvmix&&=\mns_2{+}\gamma\EEEmix{+}(1{-}\weight)p \,,\\
	  \frac{1{+}\ell}{\ell}4\nu \weight\DD^d(\vvmix) &= \TT_1^d\,,\hspace{4cm}
	 \frac{1{+}\ell}{\ell}4\nu(1{-}\weight)\DD^d(\vvmix) &&= \TT_2^d\,,\\
   \frac{1{+}\ell}{\ell}2\kappa\nabla\vartheta &= -\frac{\qqi+\left((1-\gamma)\EEEmix -\mumix\right)(\vv_1-\vv_2)}{\vartheta}\,,\hspace{1cm}
  \frac{1{+}\ell}{\ell}2\beta\mu &&= -m \,,
\end{alignat}
\begin{align}
   \frac{1{+}\ell}{\ell}\left(4\nu\DD^d(\vvmix)\nabla\weight  + 2\left(\frac{3\lambda{+}2\nu}{3} \right)\div\vvmix\nabla\weight + 2\alpha(\vv_1{-}\vv_2)\right) = -\left(\vecI + \frac{m}{2}(\vv_1{-}\vv_2) + \nabla(\gamma\EEEmix) - p\nabla\weight + \mumixvec\right)\,.
\end{align}
\end{subequations}
The value of the Lagrange multiplier $\ell$ is found by multiplying the above relations by $\div\vv_1$, $\div\vv_2$, $\DD^d(\vv_1)$, $\DD^d(\vv_2)$, $\nabla\vartheta$, $\mu$, and $\vv_1{-}\vv_2$ respectively, and summing up these relations together. One arrives at
\be
	\frac{1{+}\ell}{\ell}2\hat{\xi} = {\xi}\ \Longrightarrow \frac{1{+}\ell}{\ell} = \frac{1}{2} \Longrightarrow \ell = -2\ .
\ee 
Inserting this value for $\ell$ into \eqref{pepa:995}, we end up with  
\begin{subequations}
\label{eq-closure-relations}
	\begin{alignat}{2}
		 \mns_1 &= -\weight p + \gamma\EEEmix + \frac{3\lambda{+}2\nu}{3}\weight\div\vvmix\,,
		 \hspace{2cm}\mns_2 &&= -(1{-}\weight)p - \gamma\EEEmix + \frac{3\lambda{+}2\nu}{3}(1{-}\weight)\div\vvmix\,,
		 \\
		\TT_1^d &= 2\nu \weight\DD^d(\vvmix)\,,
		\hspace{4.5cm}\TT_2^d &&= 2\nu (1{-}\weight)\DD^d(\vvmix)\,,\\ 
		\qqi &= -\tilde{\kappa}\nabla\vartheta - \left((1-\gamma)\EEEmix -\mumix\right)(\vv_1-\vv_2)\,, 
\hspace{1cm}m &&= -\beta\mu\,,
\end{alignat}
where we introduced $\tilde{\kappa}\eqdef\vartheta\kappa$, and, finally
 \begin{align}
\label{eq:formalism-interaction-force}
& \vecI = p\nabla\weight - \nabla(\gamma\EEEmix) -\mumixvec - 2\nu\DD^d(\vvmix)\nabla\weight - \left(\frac{3\lambda{+}2\nu}{3}\right)\div\vvmix \nabla \weight - \alpha(\vv_1{-}\vv_2)-\frac{m}{2}(\vv_1{-}\vv_2)\,.
\end{align}	
\end{subequations}
The first four equations lead to the following forms for the partial Cauchy stresses:
\begin{subequations}
\label{eq:partial-cauchy-final}
	\begin{align}
		\TT_1 & = \left(\gamma\EEEmix{-}\weight p\right)\mathbb{I} + \lambda \weight \div\vvmix\mathbb{I} + 2\nu \weight \DD(\vvmix)\ ,\\
		\TT_2 & = \left(-\gamma\EEEmix{-}(1{-}\weight)p\right)\mathbb{I} + \lambda (1{-}\weight) \div\vvmix\mathbb{I} + 2\nu(1{-}\weight)\DD(\vvmix)\,.
	\end{align}
\end{subequations}
The above expressions can be simplified more by defining 
\be
	\TT_I\eqdef\TT_1+\TT_2\,,
\ee 
which then in view of \eqref{eq:partial-cauchy-final} satisfies
\be
	\label{eq:formalism-mixture-stress}
	\TT_I = -p\mathbb{I} + \lambda\div\vvmix\,\II + 2\nu\DD(\vvmix)\,,
\ee
i.e. $\TT_I$ reduces to a Cauchy stress for a compressible viscous fluid moving with the velocity field $\vvmix$. Consequently, in view of (\ref{eq:formalism-mixture-stress}), the formula for the interaction force (\ref{eq:formalism-interaction-force}) can be rewritten as follows
\be
\label{eq:formalism-interaction-force-final}
	\vecI = -\nabla(\gamma\EEEmix) - \mumixvec - \left(\alpha+\frac{m}{2}\right)(\vv_1{-}\vv_2) - \TT_I\nabla \weight\,.
\ee
\begin{remark}
\label{remark-mixture-T}
The tensor $\TT_I$ is closely related to the usual whole-mixture Cauchy stress, defined (see, e.g. Rajagopal and Tao \cite{Rajagopal-and-Tao-1995} or Hutter and J\"ohnk~\cite{hutter.k.johnk.k:continuum}) through
\begin{align}
\label{def-whole-mixture-T}
\TT \eqdef\suma \TT_\alpha -\rho_\alpha(\vv_\alpha{-}\vv)\otimes(\vv_\alpha{-}\vv)\,,
\end{align}
where $\vv$ is the barycentric velocity introduced in \eqref{def-velocities}. A whole-mixture Cauchy stress tensor defined this way admits to rewrite the whole-mixture balance of linear momentum in the form compatible with standard single-component continuum theory. Indeed, taking the sum of \eqref{eq:framework-momentum}, and making use of the condition \eqref{eq:framework-constraints}, definition of the whole-mixture density \eqref{def-rho-c-alpha}, 
and the barycentric velocity $\vv$ \eqref{def-velocities},
yields
\begin{align}
\label{aux-whole-mixture-mom-bal}
\frac{\pa \rho\vv}{\pa t} + \div(\rho\vv\otimes\vv) = \div\left(\suma\TT_\alpha - \rho_\alpha(\vv_\alpha-\vv)\otimes(\vv_\alpha-\vv)\right) + \suma\rho_\alpha\bb_\alpha\,, 
\end{align}
In this equation, the  balanced quantity is the whole-mixture momentum $\rho\vv{=}\suma\rho_\alpha\vv_\alpha$, so from this point of view, the tensor $\TT$ defined by \eqref{def-whole-mixture-T} provides the only sensible notion of a whole-mixture Cauchy stress. Note that unlike the partial Cauchy stresses $\TT_\alpha$, the whole-mixture tensor $\TT$ is in view of \eqref{aux-whole-mixture-mom-bal} accessible by measurements, as it expresses the contact-force interaction of the mixture as a whole. 

The quantity $\TT_I$ differs from $\TT$ by the quadratic diffuse flux term $\suma\rho_\alpha(\vv_\alpha{-}\vv)\otimes(\vv_\alpha{-}\vv)$,  which can be (in the approach developed here) easily computed from the individual velocities $\vv_{\alpha}$.  
\end{remark}

\subsection{Alternative derivation from the dissipation potential}
\label{remark-dissipation-potential}
Here, we would like to point out that  the use of the approach based on \textit{maximization of rate of entropy production} (see \eqref{eq-maximization-mass}) is not critical in the derivation, see however a detailed discussion concerning the validity of (not only) these thermodynamic approaches given in Rajagopal\&Srinivasa~\cite{RajagopalSrinivasa2019}, and also in Jane\v{c}ka\&Pavelka~\cite{JaneckaPavelka2018}. Alternatively,  the irreversible part of the closure relations can be formulated with the use of the {\it dissipation potential} $\mathcal{D}$
(see e.g. Edelen\cite{Edelen1972}, van Kampen \cite{vanKampen1973} or Halphen and Nguyen\cite{halphen-nguyen-1975})
: assuming that the entropy production (divided by $\vartheta$) is given 
as a generalized product of thermodynamic fluxes $\mathcal{J}_i$ and affinities $\mathcal{A}_i$, i.e. $\xi = \sum_i \mathcal{J}_i\mathcal{A}_i$ and $\mathcal{D}$ is described in terms of the affinities $\mathcal{A}_i$ then the constitutive equations for thermodynamic fluxes are given by 
\begin{align}
\label{eq-dis-pot-Ji}
\mathcal{J}_i=\frac{\partial\mathcal D}{\partial\mathcal{A}_i}\,.
\end{align}

In our setting of binary fluid mixtures, we can consider the dissipation potential $\mathcal{D}$ of the form 
\begin{align}
\label{eq:diss-pot}
	\mathcal{D} = \frac{\hat{\xi}}{2} =  \frac{3\lambda{+}2\nu}{6}(\div\vvmix)^2+\nu |\DD^d(\vvmix) |^2 + \frac{\kappa}{2}|\nabla\vartheta|^2+\frac{\beta}{2}\mu^2 + \frac{\alpha}{2}|\vv_{1}{-}\vv_2|^2\,,
\end{align}
where
\be
	3\lambda{+}2\nu{>}0,\quad \nu{>}0,\quad\kappa{>}0,\quad\beta{>} 0,\quad\alpha{>}0\,.
\ee
Recalling the equations \eqref{eq-entropy-prod-final}, the structure of the rate of entropy production (divided by $\vartheta$) takes the form:
\begin{align}
\nonumber
\xi\eqdef \vartheta\entropyproduction &=   (\mns_1{-}\gamma\EEEmix{+}\weight p)\div\vv_1 + (\mns_2{+}\gamma\EEEmix{+}(1{-}\weight)p)\div\vv_2 + \TT^d_1:\DD^d(\vv_1) + \TT^d_2:\DD^d(\vv_2)  -m \mu  \\\label{eq-entropy-prod-final-repeat} &  -\left(\frac{\qqi+\left((1-\gamma)\EEEmix -\mumix\right)(\vv_1-\vv_2)}{\vartheta}\right)\cdot\nabla\vartheta - \left(\vecI + \frac{m}{2}(\vv_1{-}\vv_2) + \nabla(\gamma\EEEmix) - p\nabla \weight + \mumixvec  \right)\cdot(\vv_1{-}\vv_2)\\ \nonumber &= \sum_i \mathcal{J}_i\mathcal{A}_i\,.
\end{align}
Let us consider the set of affinities to be the same as for the thermodynamic approach based on \textit{maximization of rate of entropy production}, i.e. $\div\vv_1$, $\div\vv_2$, $\DD^d(\vv_1)$, $\DD^d(\vv_2)$, $\mu$, $\nabla\vartheta$, $\vv_1{-}\vv_2$. If we identify the associated terms in \eqref{eq-entropy-prod-final-repeat} as the corresponding thermodynamic fluxes $\mathcal{J}_i$, then, by postulating \eqref{eq-dis-pot-Ji} for $\mathcal{D}$ given by \eqref{eq:diss-pot}, we obtain, with the use of identities \eqref{eq-xi-derivatives}, the same set of constitutive equations as before in \eqref{eq-closure-relations}.
Note also that thanks to the 2-homogeneity of $\mathcal{D}$ the choice of $\mathcal{D}$ given by \eqref{eq:diss-pot}, yields the same dissipation as \eqref{eq:diss}, since it implies 
 $\sum_i \frac{\partial\mathcal{D}}{\partial \mathcal{A}_i}\mathcal{A}_i = 2\mathcal{D} = \hat{\xi}$.

\section{Final set of governing equations} \label{sec:7}
Let us summarize the final closed system of governing equations for a binary fluid mixture. Inserting the derived constitutive relations \eqref{eq-closure-relations} into the balance equations \eqref{eq:framework-mass}, \eqref{eq:framework-momentum} and \eqref{eq-internal-e-barycentric}, we obtain the following set of partial differential equations:
\begin{itemize}
\begin{subequations}
\label{eq-final-set-compressible}
    \item Balances of mass
\begin{align}
\label{eq-final-mb1}
\frac{\pa\rho_1}{\pa t} + \div(\rho_1\vv_1) &= -\beta{\mu}\,,\\
\label{eq-final-mb2}
\frac{\pa\rho_2}{\pa t} + \div(\rho_2\vv_2) &= \beta{\mu}\,,
\end{align}
where we recall that $\beta{>}0$ is a constant (or, more generally a positive function of $\vartheta,\rho_1,\rho_2$), and the relative chemical potential $\mu$ is a function determined by the choice of free energy: ${\mu}(\vartheta,\rho_1,\rho_2) = {\mu}_1{-}{\mu}_2 = \frac{\pa\widehat{\rho\psi}(\vartheta,\rho_1,\rho_2)}{\pa\rho_1}{-}\frac{\pa\widehat{\rho\psi}(\vartheta,\rho_1,\rho_2)}{\pa\rho_2}$.
\item Balances of linear momentum
\begin{align}
\nonumber
\frac{\pa(\rho_1\vv_1)}{\pa t} + \div(\rho_1\vv_1\otimes\vv_1) &= -\weight\nabla p + \weight\nabla(\lambda\div\vvmix) + \weight\div(2\nu\DD(\vvmix))) + \rho_1\bb_1\\
\label{eq-final-mom-bal1}&-\mumixvec - \alpha(\vv_1{-}\vv_2) - \frac{\beta{\mu}}{2}(\vv_1{+}\vv_2)\,,\\\nonumber
\frac{\pa(\rho_2\vv_2)}{\pa t} + \div(\rho_2\vv_2\otimes\vv_2) &= -(1{-}\weight)\nabla p + (1{-}\weight)\nabla(\lambda\div\vvmix) + (1{-}\weight)\div(2\nu\DD(\vvmix))) + \rho_2\bb_2\\\label{eq-final-mom-bal2} & +\mumixvec + \alpha(\vv_1{-}\vv_2) + \frac{\beta{\mu}}{2}(\vv_1{+}\vv_2)\,,
\end{align}
where $0{<}\alpha(=\hat{\alpha}(\vartheta,\rho_1,\rho_2))$ and we recall that $\vvmix{=}\weight\vv_1{+}(1{-}\weight)\vv_2$.
\item Balance of energy
\begin{align}
\nonumber
\frac{\pa(\rho\ei)}{\pa t} + \div(\rho e \vvmix) &= \div(\tilde{\kappa}\nabla\vartheta) + \rho r - p\div\vvmix - \sumad\mu_\alpha\div(\rho_\alpha(\vv_\alpha{-}\vvmix)) \\\label{eq-final-energy-bal} &+ \frac{3\lambda+2\nu}{3}(\div\vvmix)^2 + 2\nu\DD^d(\vvmix):\DD^d(\vvmix) + \alpha|\vv_1-\vv_2|^2\,,
\end{align}
which is understood as an evolution equation for the temperature $\vartheta$, obtained by expressing $\rho\ei$ as a function of $\vartheta,\rho_1,\rho_2$ using \eqref{eq-aux3}:
\begin{align}
\label{eq-energy-as-function-of-temp}
\widehat{\rho\ei}(\vartheta,\rho_1,\rho_2) = \widehat{\rho\psi}(\vartheta,\rho_1,\rho_2) - \vartheta\frac{\pa\widehat{\rho\psi}(\vartheta,\rho_1,\rho_2)}{\pa\vartheta}\,,
\end{align}
applying the chain rule, and eliminating the derivatives of $\rho_1$ and $\rho_2$ with the use of \eqref{eq-final-mb1} and \eqref{eq-final-mb2}. We do not write this explicitly.
\item{Balance of entropy}
\begin{align}
\frac{\pa(\rho\eta)}{\pa t} + \div(\rho\eta\vvmix) &= \div(\kappa\nabla\vartheta) + \frac{\rho r}{\vartheta} + \frac{1}{\vartheta}\left\{
\frac{3\lambda{+}2\nu}{3}(\div\vvmix)^2+2\nu |\DD^d(\vvmix) |^2 + \kappa|\nabla\vartheta|^2+\beta\mu^2 + \alpha|\vv_{1}{-}\vv_2|^2\right\}\,.
\end{align}
\end{subequations}
\end{itemize}

\begin{remark}
\label{remark-limiting-to-pure-states}
It is instructive to look at the above system in the limiting case, when one of the substances vanishes in the sense that either $\rho_2\rightarrow 0$ (along with $\weight\rightarrow 1$) - {\it Case 1}, or $\rho_1\rightarrow 0$ (along with $\weight\rightarrow 0$) - {\it Case 2}. Assuming, in addition, that the shear viscosity, the bulk viscosity, the heat conductivity and the parameters $\alpha$ and $\beta$ are weighted as in \eqref{param-weighting}, we get \begin{align}
\nonumber
\lambda\rightarrow \lambda_1\,,\quad
\nu\rightarrow \nu_1\,,\quad
\kappa\rightarrow \kappa_1\,,\quad
\alpha\rightarrow 0\,,\quad
\beta\rightarrow 0\,,\qquad\qquad\text{for {\it Case 1}}\,,\\
\nonumber
\lambda\rightarrow \lambda_2\,,\quad
\nu\rightarrow \nu_2\,,\quad
\kappa\rightarrow \kappa_2\,,\quad
\alpha\rightarrow 0\,,\quad
\beta\rightarrow 0\,,\qquad\qquad\text{for {\it Case 2}}\,.
\end{align}
Moreover, in view of definitions \eqref{aux03} and \eqref{def-mumixvec}, we get for both cases
\begin{align}
\EEEmix\rightarrow 0\,,\qquad\mumix\rightarrow 0\,,\qquad \mumixvec\rightarrow {\bf 0}\,.
\end{align}
Since it also follows from definitions \eqref{def-vmix} and \eqref{def-q-r-e} that 
\begin{align}
\nonumber
&\vvmix\rightarrow\vv_1\,,\qquad \rho e\rightarrow \rho_1 e_1\,,\qquad \rho\eta\rightarrow \rho_1\eta_1\,,\qquad\qquad{\text{for {\it Case 1}}}\,,\\
\nonumber
&\vvmix\rightarrow\vv_2\,,\qquad \rho e\rightarrow \rho_2 e_2\,,\qquad \rho\eta\rightarrow \rho_2\eta_2\,,\qquad\qquad{\text{for {\it Case 2}}}\,,
\end{align}
the system \eqref{eq-final-set-compressible} reduces in both cases to the standard set of balance equations for a pure single component 1 ({\it Case 1}) and component 2 ({\it Case 2}).
\end{remark}

\section{Partial Cauchy stresses in thermodynamic equilibrium }
\label{sec-partical-cauchy-equilibrium}
Let us inspect the derived formulas for partial Cauchy stresses $\TT_1$ and $\TT_2$ in thermodynamic equilibrium. Defining the equilibrium as the thermodynamic process, in which all the considered affinities vanish, and denoting the equilibrium values by a dagger superscript $^\dag$, we see from \eqref{eq:partial-cauchy-final} that 
\begin{align}
\TT_1^\dag = (\gamma\EEEmix - \weight p)^\dag\II\,,\hspace{1cm}
\TT_2^\dag = (-\gamma\EEEmix - (1-\weight)p)^\dag\II\,.
\end{align}
These simple formulae are consistent with the standard picture in the sense that they yield the equilibrium value of the mixture Cauchy stress \eqref{def-whole-mixture-T} as follows
\begin{align}
\TT^\dag = \TT_I^\dag = \TT_1^\dag+\TT_2^\dag = -p\II\,, 
\end{align}
i.e., in equilibrium, the Cauchy stress reduces to the thermodynamic pressure $p$.

It is instructive to compare the equilibrium partial Cauchy stresses with the possibly simplest mixture model - mixture of ideal monoatomic gases, for which (see Appendix)
\begin{align}
    \TT^{(IG)}_1 = -xp\II\,,\hspace{1cm}\TT^{(IG)}_2 = -(1-x)p\II\,.
\end{align}
It turns out that compatibility with the ideal mixture model is satisfied in a straightforward manner for the case when $\weight{=}x$, i.e for the model where the whole-mixture velocity is weighted by the molar fractions. Indeed, in that case the equilibrium formulae read
\begin{align}
\TT^\dag_1 = (\gamma\EEEmix-xp)^\dag\II\,,\hspace{1cm}\TT^\dag_2 = (-\gamma\EEEmix - (1{-}x)p)^\dag\II\,,
\end{align}
with
\begin{equation}
\EEEmix = \rho_1 e_1(1{-}x) - \rho_2 e_2 x = \cm x(1{-}x)(\em_1-\em_2) = 0\,,
\end{equation}
as follows from \eqref{app-ienergy-eos} since for monoatomic gases $\em_1{=}\em_2{=}\frac{3}{2}R\vartheta$.

\begin{remark}
Alternative choices of the weight function $\weight$ lead to equilibrium partial pressure formulae that, in general, need not be compatible with the ideal gas mixture model. Note that formally, compatibility in the above sense for the equilibrium partial Cauchy stresses can be ensured for another weight function $\omega$ (i.e. $\weight{=}c$, $\weight{=}\phi$) by a suitable choice of $\gamma$ factor. In particular, setting $\gamma$ as $$\gamma(\rho_1,\rho_2,\vartheta) = \frac{p(\weight-x)}{\EEEmix}\ $$ yields equilibrium partial Cauchy stresses compatible with the ideal mixture model, i.e. $\TT^\dag_1{=}-xp^\dag\II$, $\TT^\dag_2{=}-(1{-}x)p^\dag\II$. Physical interpretation of such choice is, however, problematic since the terms involving $\gamma$ do not appear in the governing balance equations (as the contribution from the partial Cauchy stresses gets cancelled out by a contribution from the interaction force).


\end{remark}

\section{Imposing the constraint $\div\vvmix = 0$}
\label{sec-incompressible}
The thermodynamical approach developed in Sections \ref{sec:4}--\ref{sec-constitutive} can be, in a straightforward manner, extended to the development of model satisfying additional ``incompressibility``-type constraint
\be
\label{eq-incompressibility}
	\div\vvmix = 0\,. 
\ee

Keeping the same energy storage mechanism as above. i.e. assuming \eqref{def-psi-compressible}, 
we proceed step by step as in Section~\ref{sec-consequences}. Employing then the constraint $\div\vvmix{=}0$, which eliminates the first terms at the right-hand side of equations \eqref{eqs-auxiliary-for-of-balances2}, one arrives at the following counterpart of \eqref{eq-entropy-bal-final}:
\begin{align}
\nonumber
	& 
    -\div\entropyfluxmix + 	\rho\entropysupply + \entropyproduction =
	-\div\left(\frac{\qqi+\left((1-\gamma)\EEEmix -\mumix\right)(\vv_1-\vv_2)}{\vartheta}\right)+\frac{\rho\ri}{\vartheta}\\\nonumber &+ \frac{1}{\vartheta}\Bigg\{  (\mns_1{-}\gamma\EEEmix)\div\vv_1 + (\mns_2{+}\gamma\EEEmix)\div\vv_2 + \TT^d_1:\DD^d(\vv_1) + \TT^d_2:\DD^d(\vv_2)  -m \mu  \\\label{eq-entropy-bal-final-incompressible} &  -\left(\frac{\qqi+\left((1-\gamma)\EEEmix -\mumix\right)(\vv_1-\vv_2)}{\vartheta}\right)\cdot\nabla\vartheta - \left(\vecI + \frac{m}{2}(\vv_1{-}\vv_2) + \nabla(\gamma\EEEmix) + \mumixvec  \right)\cdot(\vv_1{-}\vv_2)
	\Bigg\}\,.
\end{align}
which differs from \eqref{eq-entropy-bal-final} by the absence of terms involving thermodynamic pressure $p$. Identifying as in the compressible case the entropy flux and the entropy supply as
\begin{subequations}
\begin{align}
\entropyfluxmix = \frac{\qqi+((1-\gamma)\EEEmix - \mumix)(\vv_1-\vv_2)}{\vartheta}\,\qquad\text{and}\qquad
\entropysupply = \frac{\ri}{\vartheta}\,,
\end{align}
the entropy production $\xi$ then takes the form
\begin{align}
\nonumber
\xi\eqdef \vartheta\entropyproduction &=   (\mns_1{-}\gamma\EEEmix)\div\vv_1 + (\mns_2{+}\gamma\EEEmix)\div\vv_2 + \TT^d_1:\DD^d(\vv_1) + \TT^d_2:\DD^d(\vv_2)  -m \mu  \\\label{eq-entropy-prod-final-in} &  -\left(\frac{\qqi+\left((1-\gamma)\EEEmix -\mumix\right)(\vv_1-\vv_2)}{\vartheta}\right)\cdot\nabla\vartheta - \left(\vecI + \frac{m}{2}(\vv_1{-}\vv_2) + \nabla(\gamma\EEEmix) + \mumixvec  \right)\cdot(\vv_1{-}\vv_2)\,.
\end{align}
\end{subequations}
Next, we include the incompressibility constraint in the ansatz for the entropy production, which we now postulate in the following form (an incompressible counterpart of \eqref{eq:diss}):
\be
\label{eq:di:ss-incompressible}
	\hat{\xi} = 2\nu |\DD^d(\vvmix) |^2 + \kappa|\nabla\vartheta|^2+\beta\mu^2 + \alpha|\vv_{1}{-}\vv_2|^2\,,
\ee 
together with
\be
	\nu{>}0,\quad\kappa{>}0,\quad\beta{>}0,\quad\alpha{>}0\,,
\ee
which ensures the non-negativity of the entropy production and keeps the setting in accordance with the second law of thermodynamics. 

Constitutive (closure) relations are again achieved by employing the principle of maximization of entropy production, but now with an additional constraint due to the assumption \eqref{eq-incompressibility}. The constrained optimization problem associated with the same set of affinities as in \eqref{eq-maximization-mass} now takes the form 
\begin{align}
	\underset{\text{w.r.t } \div\vv_1,\div\vv_2,\DD^d(\vv_1),\DD^d(\vv_2),\nabla\vartheta,\mu,\vv_1{-}\vv_2}{\text{maximize}}\left\{ \hat{\xi} + \ell_1(\hat{\xi}-\xi) + \ell_2\div\vvmix\right\}\,.
\end{align}
Employing the identity \eqref{pepa:20b}, the necessary optimality conditions read as follows:
\begin{subequations}
\begin{alignat}{2}
\label{eq:formalism-incompressible-mns-one}
  \frac{\ell_2}{\ell_1}\weight &= \mns_1{-}\gamma\EEEmix\,,
\hspace{5cm}  \frac{\ell_2}{\ell_1}(1{-}\weight)&&=\mns_2{+}\gamma\EEEmix\,,\\
\frac{1{+}\ell_1}{\ell_1}4\nu \weight\DD^d(\vvmix)&=\TT_1^d \,,
	\hspace{4cm}  \frac{1{+}\ell_1}{\ell_1}4\nu(1{-}\weight)\DD^d(\vvmix)&&=\TT_2^d\,,\\
\frac{1{+}\ell_1}{\ell_1}2\kappa\nabla\vartheta &=-\frac{\qqi+\left((1-\gamma)\EEEmix -\mumix\right)(\vv_1{-}\vv_2)}{\vartheta}\,,
\hspace{1.2cm} \frac{1{+}\ell_1}{\ell_1}2\beta\mu&&= -m\,,
\end{alignat}
\begin{align}
\frac{1{+}\ell_1}{\ell_1}\left(4\nu\DD^d(\vvmix)\nabla\weight + 2\alpha(\vv_1{-}\vv_2)\right)+ \frac{\ell_2}{\ell_1}\nabla\weight &= 
-\left(\vecI + \frac{m}{2}(\vv_1{-}\vv_2) + \nabla(\gamma\EEEmix)  + \mumixvec\right)\,. 
\end{align}
\end{subequations}
The value of the Lagrange multiplier $\ell_1$ is found by multiplying the above relations by  $\div\vv_1$, $\div\vv_2$, $\DD^d(\vv_1)$, $\DD^d(\vv_2)$, $\nabla\vartheta$, $\mu$, and $\vv_1{-}\vv_2$, respectively, and summing these relations together, one arrives at
\be
	\frac{1{+}\ell_1}{\ell_1}2\hat{\xi} = {\xi} - \frac{\ell_2}{\ell_1}\underbrace{\left(\weight\div\vv_1{+}(1{-}\weight)\div\vv_2{+}(\vv_1{-}\vv_2)\cdot\nabla \weight\right)}_{=\div\vvmix = 0} \Longrightarrow \frac{1{+}\ell_1}{\ell_1} = \frac{1}{2} \Longrightarrow \ell_1 = -2\,.
\ee 
Summing the two expressions in  \eqref{eq:formalism-incompressible-mns-one} yields
\be
	\mns\eqdef \mns_1+\mns_2 = \frac{\ell_2}{\ell_1}\,.
\ee
So, finally, we arrive at the following closure relations
\begin{subequations}
	\begin{alignat}{2}
		 \mns_1 &= \weight\mns + \gamma\EEEmix\,,
		 \hspace{4.2cm}\mns_2 &&= (1{-}\weight)\mns - \gamma\EEEmix\,,\\
		\TT_1^d &= 2\nu \weight\DD^d(\vvmix),
		\hspace{4.2cm}\TT_2^d &&= 2\nu (1{-}\weight)\DD^d(\vvmix)\,,\\
\qqi &= -\tilde{\kappa}\nabla\vartheta - \left((1-\gamma)\EEEmix -\mumix\right)(\vv_1-\vv_2)\,,
\hspace{0.5cm}m &&= -\beta\mu\,,
\end{alignat}
with $\tilde{\kappa}\eqdef\vartheta\kappa$, and, finally
\begin{align}
		 \label{eq:formalism-interaction-force-incompressible}
\vecI &= -\mns\nabla\weight - \nabla(\gamma\EEEmix) -\mumixvec - 2\nu\DD^d(\vvmix)\nabla\weight - \left(\alpha+\frac{m}{2}\right)(\vv_1{-}\vv_2)\,,
	\end{align}
\end{subequations}
The first four equations imply the following form of the partial Cauchy stresses
\begin{subequations}
	\begin{align}
		\TT_1 & = \left(\gamma\EEEmix{+}\weight\mns\right)\mathbb{I} + 2\nu\weight \DD(\vvmix)\,,\\
		\TT_2 & = \left(-\gamma\EEEmix{+}(1{-}\weight)\mns\right)\mathbb{I} + 2\nu(1{-}\weight)\DD(\vvmix)\,.
	\end{align}
\end{subequations}
Defining as before
\be
	\TT_I\eqdef\TT_1+\TT_2\,,
\ee 
we obtain
\be
	\label{eq:formalism-mixture-stress-incompressible}
	\TT_I = \mns\mathbb{I} + 2\nu\DD(\vvmix)\,.
\ee
Note that in view of (\ref{eq:formalism-mixture-stress-incompressible}), the formula for the interaction force (\ref{eq:formalism-interaction-force-incompressible}) can be rewritten as follows
\be
\label{eq:formalism-interaction-force-incompressible2}
	\vecI = -\nabla(\gamma\EEEmix) - \mumixvec - \left(\alpha+\frac{m}{2}\right)(\vv_1{-}\vv_2) - \TT_I\nabla\weight\,,
\ee
which in this form coincides with (\ref{eq:formalism-interaction-force-final}).
Finally, when plugged into the balances of mass, momentum and energy, 
(\eqref{eq:framework-mass}, \eqref{eq:framework-momentum} and \eqref{eq-internal-e-barycentric}), 
the final form of the governing equations reads as follows:
\begin{itemize}
    \item Balances of mass
\begin{subequations}
\begin{align}
\frac{\pa\rho_1}{\pa t} + \div(\rho_1\vv_1) &= -\beta{\mu}\,,\\
\frac{\pa\rho_2}{\pa t} + \div(\rho_2\vv_2) &= \beta{\mu}\,,
\end{align}
\item Balances of linear momentum
\begin{align}
\label{eq-mom-bal1-final-incompressible}
\frac{\pa\rho_1\vv_1}{\pa t} + \div(\rho_1\vv_1\otimes\vv_1) &= \weight\nabla\mns + \weight\div(2\nu\DD(\vvmix))) + \rho_1\bb_1
-\mumixvec - \alpha(\vv_1{-}\vv_2) -\frac{\beta\mu}{2}(\vv_1{+}\vv_2)\,,\\\label{eq-mom-bal2-final} 
\frac{\pa\rho_2\vv_2}{\pa t} + \div(\rho_2\vv_2\otimes\vv_2) &= (1-\weight)\nabla\mns + (1-\weight)\div(2\nu\DD(\vvmix))) + \rho_2\bb_2 +\mumixvec + \alpha(\vv_1{-}\vv_2)  + \frac{\beta\mu}{2}(\vv_1{+}\vv_2)\,.
\end{align}
\item Balance of internal energy
\begin{align}
\label{eq-final-energy-bal-incompressible}
\frac{\pa\rho\ei}{\pa t} + \div(\rho e \vvmix) &= \div(\tilde{\kappa}\nabla\vartheta) + \rho r - \sumad\mu_\alpha\div(\rho_\alpha(\vv_\alpha{-}\vvmix)) + 2\nu\DD^d(\vvmix){:}\DD^d(\vvmix) + \alpha|\vv_1{-}\vv_2|^2\,,
\end{align}
an evolution equation for temperature $\vartheta$ in view of \eqref{eq-energy-as-function-of-temp}.
\item{Balance of entropy}
\begin{align}
\frac{\pa(\rho\eta)}{\pa t} + \div(\rho\eta\vvmix) &= \div(\kappa\nabla\vartheta) + \frac{\rho r}{\vartheta} + \frac{1}{\vartheta}\left\{2\nu |\DD^d(\vvmix) |^2 + \kappa|\nabla\vartheta|^2+\beta\mu^2 + \alpha|\vv_{1}{-}\vv_2|^2\right\}\,.
\end{align}
\end{subequations}
\end{itemize}
Note that as in the compressible case, the final set of governing equations is insensitive to the splitting parameter $\gamma$, which does not appear in the final equations.

\section{Boundary conditions} \label{sec:9}
The aim of this section is to show that the introduced class of models admits a straightforward implementation of standard boundary conditions formulated for the mixture as a whole (i.e. for a single continuum). Since we are considering one temperature for both constituents, the traditional boundary conditions such as Dirichlet, Neumann or Robin type for temperature (and/or the heat flux) can be clearly applied without any change. More interestingly, we will show that the mechanical boundary conditions formulated for the mixture as a whole determine the  mechanical boundary conditions for the individual constituents and thus, in this regard, the considered model resolves one of the principal obstacles of mixture theory. 

In particular, consider a domain $\Omega$ with boundary $\pa\Omega$ endowed with outer unit normal field $\nn$. Let us consider internal flows together with a generalized slip boundary condition as introduced in Blechta et al.\cite{blechta-2020}, now formulated in terms of the mixture quantities:
\begin{subequations}
\label{bcs-mixture}
\begin{align}
\label{bcs-nonpenetration-mix}
\vvmix\cdot\nn = 0\,,\\
\label{bcs-genslip-mix}
h(\vvmix_\tau,\ssmix) = {\bf 0}\,.
\end{align}
\end{subequations}
The first condition represents the non-penetration condition (i.e. the requirement that admissible flows are internal), the latter connects
implicitly via a (continuous and monotone) function $h$, 
the tangent component of the whole-mixture velocity $\vvmix_\tau$ with the tangent whole-mixture traction $\ssmix$ defined by
\begin{align}
\label{def-smix}
\ssmix \eqdef -(\TT\nn)_\tau\,,
\end{align}
where the whole-mixture Cauchy stress $\TT$ is given by \eqref{def-whole-mixture-T}. Note that while multiple definitions of mixture velocity $\vvmix$ are take into account in this study, we only invoke one definition of the total mixture Cauchy stress $\TT$, namely the one, which is consistent with the single-component balance of momentum for the mixture as a whole (see also Remark~\ref{remark-mixture-T}). For a two-component it is given by
\begin{align}
\label{def-Tmix}
\TT\eqdef \TT_1+\TT_2 - \sum_{\alpha=1}^2 \rho_\alpha(\vv_\alpha-\vv)\otimes(\vv_\alpha-\vv)\,,
\end{align}
where $\vv$ is the barycentric velocity (see \eqref{def-velocities}). 
Now the assumption of internal flow expressed by the non-penetration kinematic condition  \eqref{bcs-nonpenetration-mix} can be naturally extended to the individual components by postulating 
\begin{align}
\label{bcs-nonpenetration-cpomponents}
\vv_1\cdot\nn = 0\,,\hspace{1cm}\vv_2\cdot\nn = 0\,.
\end{align}
While, strictly speaking, this is an independent assumption that cannot be derived from \eqref{bcs-nonpenetration-mix}, it is its only natural extension for a multicomponent fluid. 

Concerning the slip relation for individual constituents, note that
\begin{subequations}
\begin{align}
\label{bcs-ss1}
\ss_1 \eqdef -(\TT_1\nn)_\tau = -(2\nu \weight\DD(\vvmix)\nn)_\tau = 
-\left[\weight\left(\TT + \sum_{\alpha=1}^2 \rho_\alpha (\vv_\alpha{-}\vv)\otimes(\vv_\alpha{-}\vv)\right)\nn\right]_\tau = -\weight(\TT\nn)_\tau = \weight\ssmix\,,
\end{align}
where we used $(\vv_\alpha{-}\vv){\cdot}\nn{=}0$, $\alpha{=}1,2$, due to \eqref{bcs-nonpenetration-cpomponents}. Similarly, we get 
\begin{align}
\label{bcs-ss2}
\ss_2 \eqdef -(\TT_2\nn)_\tau = -(2\nu (1{-}\weight) \DD(\vvmix)\nn)_\tau =  (1{-}\weight)\,\ssmix\,.
\end{align}
\end{subequations}
Next, we express $\ssmix$ from \eqref{bcs-ss1} and \eqref{bcs-ss2}, and use the definition of $\vvmix$, see \eqref{def-vmix}, yields, after we insert these into \eqref{bcs-genslip-mix} the following two conditions
\begin{align}
    h\left(\weight(\vv_1)_\tau+(1-\weight)(\vv_2)_\tau ,\frac{\ss_1}{\weight}\right) = {\bf 0}\,,\hspace{1cm}h\left(\weight(\vv_1)_\tau+(1-\weight)(\vv_2)_\tau,\frac{\ss_2}{1-\weight}\right) = {\bf 0}\,,
\end{align}
Together with the kinematic boundary conditions \eqref{bcs-nonpenetration-cpomponents}, these two relations, constitute the implicit (and coupled) system of boundary conditions for the given mixture. 
As a particular illustrative example of the above general approach, let us consider the Navier slip condition for the mixture:
\begin{align}
\ssmix = a (\vvmix)_\tau\,,
\end{align}
where $a>0$. Then the corresponding counterparts of \eqref{bcs-ss1} and \eqref{bcs-ss2} read:
\begin{align}
\weight(\vv_1)_\tau+(1-\weight)(\vv_2)_\tau = \frac{\ss_1}{a \weight}\,,\hspace{1cm}
\weight(\vv_1)_\tau+(1{-}\weight)(\vv_2)_\tau = \frac{\ss_2}{a(1{-}\weight)}\,.
\end{align}
These two relations determine the partial tangent tractions $\ss_1$ and $\ss_2$ based on the knowledge of $\vvmix$ (which in turn is given by $\vv_1$ and $\vv_2$).

\section{Summary and concluding remarks}
\label{sec-summary}
Using a thermodynamic framework we have developed a model for flows of heat-conducting binary fluid mixtures described in terms of individual fluid densities and velocities and the whole-mixture temperature field. The framework, and consequently the whole model, is based on two constitutive assumptions for two scalar quantities: the whole-mixture Helmholtz free energy and the whole-mixture entropy production. The latter is described  in terms of a general concept of mixture velocity, that includes, in particular, the barycentric velocity, or its counterparts, where the whole-mixture velocity is weighted by volume and molar fractions. The different variants result at different equilibrium partitioning of the whole-mixture pressure between the constituents; the molar-based weighted velocity leads to the mixture of ideal (monoatomic) gases.

While individual masses and momenta of the components are distinguished in the model, and we consider their exchange through mechanical interaction and chemical reactions, the model is determined from the knowledge of mechanical properties of the mixture {\it as a whole}. In particular, only two viscosities - bulk and shear - for the whole mixture have to be specified, and the approach then determines the stresses of the individual constituents. This at first glance trivial assumption has important consequences. In particular, all the mechanical characteristics of the model should be {\it directly accessible by measurements of the whole-mixture properties}. The whole-mixture characterization also immediately translates to the formulation of boundary conditions. Consequently, the standard difficulties associated with the interpretation of the individual Cauchy stresses, common to most mixture theories, are circumvented in the setting developed in this study. It is worth of emphasizing that this whole-mixture characterization is not made at any cost to generality of the model. By considering the viscosities as (nonlinear) functions of concentrations (volume or molar fractions) of the constituents (and the temperature), see e.g.~\eqref{param-weighting}, the developed model appears to have the capacity to describe a large range of thermo-mechanical responses. The model can find the application, for instance, in the mechanics of emulsions. 

The model involves several simplifications. The two perhaps most severe ones are the two-component nature of the model and the very simple chemical kinetics of reactions. The latter can be directly improved by considering non-linear closures (e.g. in the spirit of  Bothe\cite{bothe-2015}) that are more suitable for realistic chemical reactions. An extension to a multi-component setting is also possible and should be relatively straightforward, leading to a Maxwell-Stefan type generalization of the drag dynamics among the constituents. Apart from that, the main characteristics of the model would be unchanged, the two-component setting was chosen here for the sake of simplicity.

\section{Acknowledgments} The authors acknowledge the support of the project No. 18-12719S
financed by Czech Science Foundation (GACR). The authors are members of the Ne\v{c}as
center for mathematical modelling. 



\providecommand{\url}[1]{\texttt{#1}}
\providecommand{\urlprefix}{}
\providecommand{\foreignlanguage}[2]{#2}
\providecommand{\Capitalize}[1]{\uppercase{#1}}
\providecommand{\capitalize}[1]{\expandafter\Capitalize#1}
\providecommand{\bibliographycite}[1]{\cite{#1}}
\providecommand{\bbland}{and}
\providecommand{\bblchap}{chap.}
\providecommand{\bblchapter}{chapter}
\providecommand{\bbletal}{et~al.}
\providecommand{\bbleditors}{editors}
\providecommand{\bbleds}{eds: }
\providecommand{\bbleditor}{editor}
\providecommand{\bbled}{ed.}
\providecommand{\bbledition}{edition}
\providecommand{\bbledn}{ed.}
\providecommand{\bbleidp}{page}
\providecommand{\bbleidpp}{pages}
\providecommand{\bblerratum}{erratum}
\providecommand{\bblin}{in}
\providecommand{\bblmthesis}{Master's thesis}
\providecommand{\bblno}{no.}
\providecommand{\bblnumber}{number}
\providecommand{\bblof}{of}
\providecommand{\bblpage}{page}
\providecommand{\bblpages}{pages}
\providecommand{\bblp}{p}
\providecommand{\bblphdthesis}{Ph.D. thesis}
\providecommand{\bblpp}{pp}
\providecommand{\bbltechrep}{}
\providecommand{\bbltechreport}{Technical Report}
\providecommand{\bblvolume}{volume}
\providecommand{\bblvol}{Vol.}
\providecommand{\bbljan}{January}
\providecommand{\bblfeb}{February}
\providecommand{\bblmar}{March}
\providecommand{\bblapr}{April}
\providecommand{\bblmay}{May}
\providecommand{\bbljun}{June}
\providecommand{\bbljul}{July}
\providecommand{\bblaug}{August}
\providecommand{\bblsep}{September}
\providecommand{\bbloct}{October}
\providecommand{\bblnov}{November}
\providecommand{\bbldec}{December}
\providecommand{\bblfirst}{First}
\providecommand{\bblfirsto}{1st}
\providecommand{\bblsecond}{Second}
\providecommand{\bblsecondo}{2nd}
\providecommand{\bblthird}{Third}
\providecommand{\bblthirdo}{3rd}
\providecommand{\bblfourth}{Fourth}
\providecommand{\bblfourtho}{4th}
\providecommand{\bblfifth}{Fifth}
\providecommand{\bblfiftho}{5th}
\providecommand{\bblst}{st}
\providecommand{\bblnd}{nd}
\providecommand{\bblrd}{rd}
\providecommand{\bblth}{th}

\appendix
\section{Mixture of ideal gases} \label{sec-ideal-gas}

In an $N$-component mixture of ideal gasses with common temperature, each component behaves as an ideal gas, in particular we have the following state equations (see e.g. \cite{callen.hb:thermodynamics})
\begin{itemize}
\item Partial pressures:
\begin{align}
\label{eq:ideal-mix-p-alpha}	
	p_\alpha = \hat{p}_\alpha(\vartheta,\rho_\alpha) = \frac{R\vartheta}{\mathrm{M}_\alpha}\rho_\alpha\,,\hspace{1cm}\alpha=1,\dots,N\,,
\end{align}
where $R$ is the universal gas constant ($R=8.3144598$ J K$^{-1}$ mol$^{-1}$).
\item
Specific internal energy
\begin{align}
\label{app-ienergy-eos}
e_\alpha = \bar{e}_\alpha(\vartheta)= z_\alpha\frac{R\vartheta}{\mathrm{M}_\alpha} ,\hspace{1cm}\alpha=1,\dots,N\,,
\end{align}
where $z_\alpha$ is the ``equi-partitioning'' term (e.g. $\frac{3}{2}$ for a monoatomic gas).
\item Entropy
\be
\eta_\alpha = \hat{\eta}_\alpha(\vartheta,\rho_\alpha)=z_\alpha\frac{R}{\mathrm{M}_\alpha}\ln\vartheta-\frac{R}{\mathrm{M}_\alpha}\ln\rho_\alpha +d_\alpha\ ,\hspace{1cm}\alpha=1,\dots,N\ ,
\ee
where $d_\alpha$ are constants,
\end{itemize}
the mixture energy and entropy are
\begin{align}
\rho e = \suma \rho_\alpha e_\alpha\,,\hspace{1cm}\rho\eta = \suma \rho_\alpha\eta_\alpha\,,
\end{align}
and thus, the fundamental thermodynamic relation can be constructed from the above state equations for $e_\alpha$, $\eta_\alpha$, $p_\alpha$ in for Helmholtz free energy
\begin{align}
\label{eq:chem-Helmholtz}
\widehat{\rho\psi}(\vartheta,\rho_1,\dots,\rho_N)=\suma \rho_\alpha\psi_\alpha =\suma \rho_\alpha(\hat{e}_\alpha(\vartheta)-\vartheta\hat{\eta}_\alpha(\vartheta,\rho_\alpha))=
\suma \rho_\alpha\left(z_\alpha \frac{R\vartheta}{\mathrm{M}_\alpha}\right)-\rho_\alpha\vartheta\left(z_\alpha\frac{R}{\mathrm{M}_\alpha}\ln\vartheta-\frac{R}{\mathrm{M}_\alpha}\ln\rho_\alpha + d_\alpha\right)\,.
\end{align}
The chemical potential $\mu_\alpha$ defined by \eqref{def-chem-pot} then reads
\begin{align}
\nonumber
\mu_\alpha = \left.\frac{\partial{\widehat{\rho\psi}(\vartheta,\rho_1,\dots,\rho_N)}}{\partial\rho_\alpha}\right|_{\vartheta,\rho_{\beta\neq\alpha}}=&
\underbrace{\left(z_\alpha\frac{R\vartheta}{\mathrm{M}_\alpha}+\beta_\alpha\right)}
_{\hat{e}_\alpha(\vartheta)}-
\vartheta\underbrace{\left(z^\alpha\frac{R\vartheta}{\mathrm{M}_\alpha}\ln\vartheta-\frac{R}{\mathrm{M}_\alpha}\ln\rho_\alpha+ d_\alpha\right)}
_{\hat{\eta}_\alpha(\vartheta,\rho_\alpha)}+\underbrace{\frac{R\vartheta}{\mathrm{M}_\alpha}}_{\frac{\hat{p}_\alpha(\vartheta,\rho_\alpha)}{\rho_\alpha}}
\\\label{aux-a1}
 =& \hat{e}_\alpha(\vartheta) - \vartheta\hat{\eta}_\alpha(\vartheta,\rho_\alpha)+ \frac{\hat{p}_\alpha(\vartheta,\rho_\alpha)}{\rho_\alpha}\,.
\end{align}
Consequently, the mixture thermodynamic pressure $p$, defined by \eqref{def-pressure}, reads
\begin{align}
p = -\rho\ei + \vartheta\rho\eta + \suma\rho_\alpha\mu_\alpha = R\vartheta\suma\frac{\rho_\alpha}{M_\alpha}\,.
\end{align}
As a result, we see that
\begin{align}
\label{eq-pressure-formula}
p = \suma p_\alpha\,,\hspace{1cm}\text{and}\hspace{1cm}p_\alpha = p \frac{ \frac{\rho_\alpha}{M_\alpha}}{\sum_{\beta=1}^N\frac{\rho_\beta}{M_\beta}} = p \frac{\cm_\alpha}{\cm} = p x_\alpha\,,
\end{align}
so that Dalton's law for ideal mixtures is compatible with the definition \eqref{def-pressure}, and the relation of partial and total mixture pressure is via molar fractions $x_\alpha$. Finally, using $\rho_\alpha = \frac{M_\alpha}{R\vartheta} x_\alpha p$ in \eqref{aux-a1}, allows to rewrite the chemical potential in the standard form for ideal mixtures:
\begin{align}
\mu_\alpha = \mu_\alpha^0(\vartheta,p) + \frac{R\vartheta}{M_\alpha}\ln x_\alpha\,,\qquad\text{where}\qquad \mu^0_\alpha(\vartheta,p) \eqdef \hat{e}_\alpha(\vartheta) - \vartheta\hat{\eta}_\alpha\left(\vartheta,\frac{M_\alpha p}{R\vartheta}\right)+ \frac{R\vartheta}{M_\alpha}\,.
\end{align}

\end{document}

%% file: packages.tex
\usepackage{latexsym,amsmath,amsthm,amsfonts}
\usepackage{graphicx,epsfig} 
\usepackage{amsmath}
\usepackage{amsxtra}
\usepackage{accents}
\usepackage{amstext}
\usepackage{latexsym}
\usepackage{dsfont}
\usepackage[T1]{fontenc}
\usepackage{extarrows}
\usepackage{graphicx}
\usepackage{amsthm}
\usepackage{subfig}
\usepackage{enumerate}
\usepackage{color}
\usepackage{float}
\usepackage{thmtools}
\usepackage{comment}
\usepackage{mathtools}
\usepackage[most]{tcolorbox}

%% file: macros-generic.tex
\newcommand{\OS}[1]{{\color{blue}{#1}}}
\newcommand{\REPLACE}[2]{{\color{red}\sout{#1}}{\ \color{black}\uline{#2}}}
\newcommand{\MAYBE}[2]{{\color{blue}{#1}}{{\color{blue}{/#2}}}}
\newcommand{\INSERT}[1]{{\color{blue}\uuline{#1}\color{black}}}
\newcommand{\DELETE}[1]{{\color{red}\sout{#1}\color{black}}\ }
\newcommand{\CHECK}[1]{{\color{red}{#1}}\ }
\newcommand{\COLOR}[2]{{\color{#1}{#2}}}
\newcommand{\REM}[1]{\marginpar{{\color{red}#1}}}

\renewcommand{\div}{\mathop{\rm div}\nolimits}
\newcommand{\tr}{\mathop{\rm tr}\nolimits}
\def\entropyflux{\tzrg{\Phi}}
\def\entropyfluxmix{\tzrg{\Phi}
^{\mbox{\tiny{mix}}}}

\def\entropyfluxm{\tzrg{\Phi}^{\mbox{\tiny{M}}}}
\def\entropysupply{h}
\def\entropyproduction{\Pi}
\def\cm{c^{\mbox{\tiny{M}}}}
\def\mf{x}
\def\mns{\pi}

\newcommand{\bb}{{\bf{b}}}

\newcommand{\qq}{{\bf{q}}}
\newcommand{\vv}{{\bf v}}
\newcommand{\vvm}{{\bf v}^{\mbox{\tiny{M}}}}
\newcommand{\vvphi}{{\bf v}^{{\tiny{\phi}}}}

\newcommand{\vvmix}{{\bf v}^{\mbox{\tiny{mix}}}}
\newcommand{\uu}{{\bf u}}
\newcommand{\uum}{{\bf u}^{\mbox{\tiny{M}}}}

\newcommand{\jj}{{\bf j}}
\newcommand{\jjm}{{\bf j}^{\mbox{\tiny{M}}}}
\newcommand{\jjmix}{{\bf j}^{\mbox{\tiny{mix}}}}

\newcommand*{\dotm}[1]{
  \accentset{\bigcirc}{#1}}

\renewcommand*{\dot}[1]{
  \accentset{\mbox{\large\bfseries .}}{#1}}

\renewcommand{\aa}{{\bf a}}
\def\mum{\mu^{\mbox{\tiny{M}}}}

\newcommand{\DD}{\mathbb{D}}

\newcommand{\ei}{e}
\newcommand{\qqi}{{\bf{q}}}
\newcommand{\ri}{r}

\newcommand{\EEE}{E_{12}}
\newcommand{\EEEmix}{E_{12}^{\mbox{\tiny{mix}}}}
\newcommand{\mumix}{\mu_{12}^{\mbox{\tiny{mix}}}}

\newcommand{\mumixvec}{\boldsymbol{\mu}_{12}^{\mbox{\tiny{mix}}}}

\newcommand{\EEEm}{E_{12}^{\mbox{\tiny{M}}}}

\newcommand{\om}{_\Omega}
\newcommand{\tzrg}[1]{\mbox{\boldmath $#1$}}
\newcommand{\supply}{\mathcal{S}}
\newcommand{\production}{\xi}
\newcommand{\inter}{\Pi}
\newcommand{\vecI}{{\mathbf{I}}}
\renewcommand{\ss}{\mathbf{s}}
\newcommand{\ssmix}{\mathbf{s}^{mix}}

\newcommand{\bea}{\begin{eqnarray}}
\newcommand{\eea}{\end{eqnarray}}
\newcommand{\be}{\begin{equation}}
\newcommand{\ee}{\end{equation}}
\newcommand{\pa}{\partial}
\renewcommand{\SS}{\tzrg{\sigma}}
\newcommand{\stressc}{\tzrg{\mathrm{T}_c}}
\newcommand{\grad}[1]{\mathrm{grad} #1}
\newcommand{\Grad}[1]{\mathrm{Grad} #1}
\newcommand{\heatflux}{{\bf q}}
\newcommand{\heatfluxtot}{{\bf q}^{tot}}
\newcommand{\WW}{\mathbb{W}}
\newcommand{\FF}{\mathbb{F}}
\newcommand{\CC}{\mathbb{C}}
\newcommand{\cc}{{\bf c}}
\newcommand{\BB}{\mathbb{B}}
\newcommand{\TT}{\mathbb{T}}
\newcommand{\TTmix}{\mathbb{T}^{mix}}
\newcommand{\KK}{\mathbb{K}}
\newcommand{\II}{\mathbb{I}}
\newcommand{\VV}{\mathbb{V}}

\newcommand{\ww}{{\bf w}}

\newcommand{\muc}{{\tzrg{\mu}_c}}
\newcommand{\mucc}{\mu_c}

\newcommand{\xich}{\xi^\mathrm{CH}}

\newcommand{\LL}{\mathbb{L}}
\newcommand{\bm}{\mathbf}

\newcommand{\eqdef}{\stackrel{\mathrm{def}}{=}}

\newcommand{\nn}{\tzrg{n}}

\newcommand{\weight}{\omega}
\newcommand{\rhot}{\rho^{\rm m}}
\newcommand{\mug}{{\tzrg\mu}^{\mathrm{grad}}}
\newcommand{\rr}{\mathrm{r}^*}
\newcommand{\mutot}{\mu^{tot}}

\newcommand{\xx}{{\bf x}}
\newcommand{\xix}{{\tzrg{\xi}}}

\newcommand{\QQ}{{\bf Q}}

\newcommand{\yy}{{\bf y}}
\newcommand{\hh}{{\bf h}}
\newcommand{\XX}{{\bf X}}

\newcommand{\ssigma}{s}

\newcommand{\iforce}{{\bf I}}

\newcommand{\spin}{{\bf s}}
\newcommand{\cs}{\mathbb{M}}
\newcommand{\intercouptensor}{\mathbb{P}}
\newcommand{\spinsup}{{\bf l}}
\newcommand{\intercoup}{{\bf p}}
\newcommand{\stressa}{{\bf A}}
\newcommand{\suma}{\sum_{\alpha=1}^N}
\newcommand{\sumad}{\sum_{\alpha=1}^2}
\newcommand{\suman}{\sum_{\alpha=1}^{N-1}}
\newcommand{\sumb}{\sum_{\beta=1}^N}
\newcommand{\sumbn}{\sum_{\beta=1}^{N-1}}

\newcommand{\vecko}{\vec{v}}
\newcommand{\becko}{\vec{B}}
\newcommand{\A}{\vec{A}}
\newcommand{\beckokr}{\vec{B}_{k_R}}
\newcommand{\beckokpt}{\vec{B}_{k_{p(t)}}}
\newcommand{\beckokptj}{\vec{B}_{k_{p_1(t)}}}
\newcommand{\beckokptd}{\vec{B}_{k_{p_2(t)}}}
\newcommand{\ceckokptj}{\vec{C}_{k_{p_1(t)}}}
\newcommand{\ceckokptd}{\vec{C}_{k_{p_2(t)}}}
\newcommand{\deckokptj}{\vec{D}_{k_{p_1(t)}}}
\newcommand{\deckokptd}{\vec{D}_{k_{p_2(t)}}}
\newcommand{\efkptj}{\vec{F}_{k_{p_1(t)}}}
\newcommand{\efkptd}{\vec{F}_{k_{p_2(t)}}} 
\newcommand{\qecko}{\vec{q}}
\newcommand{\decko}{\vec{D}}
\newcommand{\acko}{\vec{A}}
\newcommand{\icko}{\vec{I}}
\newcommand{\elko}{\vec{L}}
\newcommand{\esko}{\vec{S}}
\newcommand{\eskokol}{\stackrel{\circ}{\vec{S}}}
\newcommand{\beckool}{\stackrel{\triangledown}{\vec{B}}}
\newcommand{\ackool}{\stackrel{\triangledown}{\vec{A}}}
\newcommand{\eskotriangle}{\stackrel{\triangledown}{\vec{S}}}
\newcommand{\ddp}{{\vec{D}}^{\delta}}
\newcommand{\tdp}{{\vec{T}}^{\delta}}
\newcommand{\bdp}{{\vec{B}}^{\delta}}
\newcommand{\bddp}{\stackrel{\triangledown}{\vec{B}}\hspace{-3pt}^{\delta}}
\newcommand{\bmdp}{{\mathcal{B}}^{\delta}}
\newcommand{\bmddp}{\stackrel{\triangledown}{\mathcal{B}}\hspace{-3pt}^{\delta}}
\newcommand{\tauold}{\stackrel{\triangledown}{\vec{\tau}}}
\newcommand{\fpt}[1]{\vec{F}_{p_1(t)}^{\rm #1}}
\newcommand{\pder}[2]{\frac{\partial #1}{\partial #2}}
\newcommand{\pdert}[1]{\frac{\partial #1}{\partial t}}
\newcommand{\Div}{\mathop{\rm Div}\nolimits}

\renewcommand{\SS}{\tzrg{\sigma}}
\newcommand{\stress}{\mathbb{T}}
\newcommand{\Surface}{\Gamma}
\newcommand{\ga}{_\Surface}
\newcommand{\pga}{_{\pa\Surface}}

\newcommand{\ienergy}{e}
\renewcommand\em{e^{\mbox{\tiny{M}}}}
\newcommand{\psii}{\psi_I}
\newcommand{\etam}{\eta^{\mbox{\tiny{M}}}}
\font\zpd=pzdr at 16pt
\def\ding#1{{\zpd \char#1}}
\def\arrow{\ding{226}}
\def\handpen{\ding{45}}
\def\hand{\ding{43}}
\def\dx{dx}
\def\ds{dS}
\def\dl{dl}
\def\tflux{\mathcal{F}}
\def\tprodu{\mathcal{P}}
\def\tsupply{\mathcal{S}}
\def\flux{\Phi}
\def\supply{\Sigma}

\def\vvt{{\bf v}^\|\ga}
\def\vvp{\mathrm{v}^\perp\ga}

\def\vvta{{{\bf v}_\alpha^\|}\ga}
\def\vvpa{{\mathrm{v}_\alpha^\perp}\ga}

\def\tens#1{\boldsymbol{\mathsf{#1}}}
\def\pp#1#2{\frac{\partial #1}{\partial #2}}
\def\energy{\ensuremath{e}}

\def\fmol{y^{\mbox{\tiny{M}}}}
\def\etamol{\eta^{\mbox{\tiny{M}}}}
\def\etamolb{\bar{\eta}^{\mbox{\tiny{M}}}}
\def\mumolb{\bar{\mu}^{\mbox{\tiny{M}}}}

\def\mumolv{{\tzrg{\mu}^{\mbox{\tiny{M}}}}}


\newcommand{\error}[1]
{
\begin{center}
\bf #1
\end{center}
}